\documentclass[12pt]{amsart}

\usepackage{amssymb,amsfonts,latexsym,amscd,epsfig,psfrag}

\begin{document}

\def\A{{\mathcal A}}
\def\alg{{\mathfrak A}}
\def\amp{{\rm Amp}[\pi]}
\def\ampl{\rm Amp}

\def\Bound{{\mathcal B}}

\def\bra{\Big\langle}

\def\C{{\Bbb C}}
\def\card{{\rm card}}
\def\cF{{\mathcal F}}
\def\cL{{\mathcal L}}
\def\com{{c_V}}
\def\cT{{\mathcal T}}
\def\cV{{\mathcal V}}
\def\Cnm{\Big(\begin{array}{c}2n\\2l_1\dots 2l_m\end{array}\Big)}
\def\Cbnm{\Big(\begin{array}{c}2\bar n\\2l_1\dots 2l_m\end{array}\Big)}
\def\CSnm{\Big(\begin{array}{c}2 n\\|S_1|\dots |S_m|\end{array}\Big)}
\def\CSnn{\Big(\begin{array}{c}2 n\\|S_1|\dots|S_{n-1}|\end{array}\Big)}
\def\CbSnm{\Big(\begin{array}{c}2\bar n\\|S_1|\dots|S_m|\end{array}\Big)}
\def\CbSnn{\Big(\begin{array}{c}2\bar n\\|S_1|\dots |S_{\bar n-1}|\end{array}\Big)}
\def\cE{{\mathcal E}}
\def\crit{{\mathcal C}}

\def\Dell{{\alg}_{L,\e,\delta,\ell}^{(\omega)}}
\def\cDell{\alg_L\setminus \alg_{L,\e,\delta,\ell}^{(\omega) }}
\def\Dom{\mathfrak{Dom}}
\def\dist{{\rm dist}}
\def\dup{d\up }
\def\dupm{d\up^{(m+1)}}
\def\dupn{d\up^{(n+1)}}
\def\dvpm{d\uv^{(m+1)}}
\def\dvpn{d\vp^{(n+1)}}

\def\ellipse{{\rm ell}}
\def\Exp{{\Bbb E}}
\def\e{\varepsilon}
\def\en{{e_{\Delta}}}
\def\esplit{\end{split}}
\def\etwo{{e_{2D}}}

\def\Fou{{\mathcal F}}

\def\gex{\frac{1}{5}}
\def\gexc{\frac{4}{5}}

\def\H{{\mathcal H}}
\def\Hpl{{\Bbb H}}

\def\Im{{\rm Im}}

\def\ket{\Big\rangle}

\def\Lap{\Delta}
\def\lb{\left[}
\def\lev{{\mathcal L}}
\def\LL{\Lambda_L}
\def\LLinv{\frac{1}{|\LL|}}
\def\LLs{\Lambda_L^*}
\def\LLsqinv{\frac{1}{|\LL|^{2n+2}}}

\def\mes{{\rm mes}}

\def\N{{\Bbb N}}

\def\nm{{|\!|\!|\,}}

\def\pip{\tau}
\def\pvar{u}

\def\qm{q^{(m+1)}}

\def\R{{\Bbb R}}
\def\Re{{\rm Re}}
\def\rb{\right]}

\def\rc{\frac{1}{2}}
\def\Rem{R}

\def\Sc{{\mathcal S}}
\def\split{\begin{split}}
\def\summn{\sum_{0\leq m\in n-2\N_0}}
\def\summnp{\sum_{0\leq m\in\min\{n,n'\}-2\N_0}}
\def\Sym{{\mathfrak S}}

\def\Tor{\Bbb T}
\def\tk{\tilde k}
\def\tt{\theta}
\def\Th{\Xi}
\def\th{q'}
\def\Thinf{\Th }
\def\tqm{\tilde q^{(m+1)}}
\def\tLL{\tilde\Lambda_L}

\def\ua{\underline{a}}
\def\up{\underline{p}}
\def\uq{\underline{q}}
\def\uk{\underline{k}}
\def\un{\underline{n}}
\def\ux{\underline{x}}
\def\uv{\underline{v}}
\def\uw{\underline{w}}
\def\utk{\underline{\tilde\vk}}
\def\uvw{\underline{w}}

\def\Val{{\rm Val}}
\def\vac{\Omega_f}

\def\xif{\frac{\xi}{2}}

\def\Z{{\Bbb Z}}

\def\vx{{ x}}
\def\vy{{ y}}
\def\ve{{ e}}
\def\vk{{ k}}
\def\vl{{ l}}
\def\vm{{ m}}
\def\vn{{ n}}
\def\vp{{ p}}
\def\vQ{{ Q}}
\def\vq{{ q}}
\def\vr{{ r}}
\def\vv{{ v}}
\def\vw{{ w}}

\def\Hspace{{\mathfrak H}}
\def\Mspace{{\mathfrak M}}
\def\Polyd{{\mathfrak V}}
\def\Wspace{{\mathfrak W}}
\def\Uspace{{\mathfrak U}}
\def\Tspace{{\mathfrak T}}
\def\Pl{{\mathcal P}}

\def\1{{\bf 1}}

\def\eqnn{\begin{eqnarray*}}
\def\eeqnn{\end{eqnarray*}}
\def\eqn{\begin{eqnarray}}
\def\eeqn{\end{eqnarray}}
\def\bal{\begin{align}}
\def\eal{\end{align}}

\def\prf{\begin{proof}}
\def\endprf{\end{proof}}

\newtheorem{theorem}{Theorem}[section]
\newtheorem{definition}{Definition}[section]
\newtheorem{proposition}{Proposition}[section]
\newtheorem{lemma}{Lemma}[section]

\title{ Localization lengths and Boltzmann limit for the
Anderson model at small disorders in dimension 3}

\author{Thomas Chen}

\address{Department of Mathematics,
Princeton University,
807 Fine Hall, Washington Road,
Princeton, NJ 08544, U.S.A.}

\email{tc@math.princeton.edu}

\date{}
\maketitle

\begin{abstract}
We prove lower bounds on the localization length of eigenfunctions
in the three-dimensional Anderson model at weak disorders. Our results
are similar to those obtained by Schlag, Shubin and Wolff,
\cite{shscwo},
for dimensions one and two. We prove that with probability one, most
eigenfunctions have localization
lengths bounded from below by
$O(\frac{\lambda^{-2}}{\log\frac1\lambda})$,
where $\lambda$ is the disorder strength. This is
achieved by time-dependent methods which generalize those developed by
Erd\"os and Yau
\cite{erdyau} to the lattice and non-Gaussian case.
In addition, we show that the macroscopic limit of the corresponding
lattice random Schr\"odinger dynamics is governed by a linear
Boltzmann equation.
\end{abstract}

\section{Introduction}
The Anderson model in dimension $d$ is defined by the
discrete random Schr\"odinger operator
$$
        (H_\omega\psi)(x)=-\frac12(\Lap \psi)(x) +
        \lambda \omega(x)\psi(x) \;,
$$
acting on $\ell^2(\Z^d)$, where  $\lambda$ is a small coupling
constant, accounting for the strength of the disorder.
$$
        (\Lap \psi)(x):=2d\psi(\vx)-\sum_{|x-y|=1}\psi(y)
$$
is the nearest neighbor lattice Laplacian, and $\omega(x)$ shall,
for $x\in\Z^d$, be bounded, i.i.d. random variables.
In the present paper, we study the case $d=3$, and prove that
with probability one, most eigenfunctions of $H_\omega$ have
localization lengths bounded from below by
$O(\frac{\lambda^{-2}}{\log\frac1\lambda})$.
In contrast to $d=1,2$, we note that there are no restrictions
on the energy range for the validity of this result.
Furthermore, we derive the macroscopic limit of the
quantum dynamics in this system, and prove that
it is governed by the linear Boltzmann equations.

The present paper is closely related to work of
L. Erd\"os and H.-T. Yau, \cite{erdyau}, in which the
weak coupling and
hydrodynamic limit has been derived for a random Schr\"odinger
equation in the continuum $\R^d$, $d=2,3$,
for a Gaussian random potential.
For macroscopic time and space variables $(T,X)$,
microscopic variables $(t,x)$, and the scaling
$(X,T)=\lambda^{2}(x,t)$, where $\lambda$ is the coupling
constant in the continuum analogue of $H_\omega$,
these authors established in the limit $\lambda\rightarrow0$
that the macroscopic dynamics is governed by a
{\em linear Boltzmann equation},
and thus ballistic, globally in $T>0$.
We note that the corresponding local in $T>0$ result
was first proved by H. Spohn \cite{sp}.
For a time scale larger than $O(\lambda^{-2})$,
L. Erd\"os, M. Salmhofer and H.-T. Yau
have very recently succeeded in establishing that the macroscopic
dynamics in $d=3$ is determined by a diffusion equation,
\cite{erdsalmyau}.

The problem addressed in the present paper is,
on the other hand, closely related to recent work of
W. Schlag, C. Shubin and T. Wolff, \cite{shscwo}.
Based on techniques of harmonic analysis,
it was established in \cite{shscwo}
for the Anderson model at small disorders in $d=1,2$
that with probability one, most eigenstates are
in frequency space concentrated on shells of thickness
$\leq\lambda^{2}$ in $d=1$, and $\leq\lambda^{2-\delta}$ in $d=2$.
The eigenenergies are required to be bounded away from the edges of the
spectrum of $-\frac12\Delta_{\Z^d}$, and in $d=2$,
also away from its center. By the uncertainty principle,
this implies lower bounds of order $O(\lambda^{-2})$ in $d=1$,  
and $O(\lambda^{-2+\delta})$ in $d=2$,
on the localization lengths in position
space. Closely related to their  work are
the papers \cite{mapori1,mapori2}
by J. Magnen, G. Poirot, V. Rivasseau, and
\cite{po} by G. Poirot, which address properties of the
Greens functions associated to $H_\omega$.

The proof the main results in the present paper
uses an extension of the time-dependent techniques
of L. Erd\"os and H.-T. Yau  in \cite{erdyau}
to the lattice, and to non-Gaussian random potentials.
Higher correlations, which are now abundant,
are shown to have an insignificant effect, hence the character
of our results does not differ from that obtained in the Gaussian case.
Furthermore, bounds on the amplitudes of certain Feynman diagrams of
"crossing" structure are much harder to obtain in the lattice
than in the continuum model, due to the significantly more
complicated geometry of energy level surfaces.
We have adapted part of our notation and nomenclature to
\cite{erdyau}, in order to
facilitate the referencing of results.

The link between the lower bounds on the localization lengths of
eigenfunctions,
and the Schr\"odinger dynamics generated by $H_\omega$
is a joint result with L. Erd\"os and H.-T. Yau  included in
this paper.
The author is deeply grateful to them for their support
and generosity.

\section{Definition of the model and statement of the main theorem}

We consider the discrete random Schr\"odinger operator
\eqn
         H_\omega = -\frac12\Lap  + \lambda V_\omega \;
\eeqn
acting on $\psi\in\ell^2(\Z^3)$.
The impurity potential is given by
\eqn
        V_{\omega}(\vx) =\sum_{ \vy \in\Z^3 } \omega_\vy \delta(\vx-\vy)
\;,
\eeqn
where $\omega_y$ are bounded, independent, identically distributed
random variables, of mean 0, and normalized variance.
For each $x\in\Z^3$, $\omega_x$ is a random variable
on a single site probability space $(J,F,\mu)$, where $J$ is a Borel subset
of $\R$ with
$|J|:=\sup_{\omega,\omega'\in J}|\omega-\omega'|<\infty$,
$F$ is the $\sigma$-algebra of Borel subsets of $J$, and
$\mu$ is a probability measure
on $F$.
$V_\omega$ is a random field over $\Z^3$
realized on the probability space
$(\Omega,{\mathcal F},{\Bbb P})$, with $\Omega=\times_{\Z^3}J$, where
${\mathcal F}$ is the $\sigma$-algebra
generated by the cylinder sets induced by $F$, and the probability
measure ${\Bbb P}$ is given by $\times_{\Z^3}\mu$.
For simplicity, we assume $\mu$ to be even, $\mu(I)=\mu(-I)$,
for all  $I\in F$.
Then, $\Exp[\omega_\vx^{2m+1}]=0$
$\forall x\in\Z^3$, $\forall\, m\geq0$.
This reduces some of the notation in our
analysis, but  for our methods to apply, only $\Exp[\omega_x]=0$
is necessary. Clearly, $\Exp[\omega_\vx^{2m}]<|J|^{2m}$ for all $m$, but we
shall here use the moment bounds
\eqn\label{omcorrdef}
        \Exp[\omega_\vx^{2m}] =: \tilde c_{2m} \leq (2m)! \, \com
        \; \; , \; \;
        \tilde c_2=1
         \; , \; \forall\vx\in\Z^3 \; , \; \forall m\geq1 \;,
\eeqn
for a constant $\com<\infty$ which is independent of $m$ and $|J|$.
This allows for a generalization of our results to cases of
unbounded random variables, which we expect to be straightforward.
We shall here not further discuss the latter issue.

We use the convention
\eqn
            \hat f(\vk) &\equiv&\Fou(f)(\vk)= \sum_{\vx\in\Z^3}
            e^{-2\pi i\vk\cdot\vx} f(\vx)
            \nonumber\\
            \check g(\vx)
            &\equiv&\Fou^{-1}(g)(\vx)= \int_{\Tor^3} dk\; g(\vk) e^{2 \pi i \vk\cdot\vx} \;
\eeqn
for the Fourier transform and its inverse.
Then,
\eqn
        (\Lap  f)\hat{\;}(\vk) &=&  -\, 2\,\en(\vk) \hat f(\vk) \;,
        \nonumber\\
        \en(\vk) &:=& \sum_{i=1}^3 \big( 1- \cos(2\pi k_i) \big)
        =2\sum_{i=1}^3 \sin^2(\pi k_i)
        \label{kinendef}
\eeqn
is the expression for the kinetic energy in frequency space.

Let  $L\gg\lambda^{-2}$, and
$\Lambda_L:=[-L,L]^3\cap\Z^3$.
For $m\in\N_0$ and $\ell\in\R$ with $m\leq\ell\ll L$, let
\eqn
        h_\ell(m) &:=& \left\{\begin{array}{ll}1&\;\;{\rm if}\; 0\leq m\leq
        \frac{\lfloor\ell\rfloor}{2}\\
        2-\frac{2m}{\lfloor\ell\rfloor} &\;\;
        {\rm if}\;\frac{\lfloor\ell\rfloor}{2}< m
        \leq
        \lfloor\ell\rfloor
        \\
        0&\;\;{\rm otherwise}\;\end{array}\right.
        \nonumber\\
        K_\ell(x)&:=& \prod_{j=1}^3  h_\ell(|x_j|)  
        \nonumber\\
        R_{x,\delta, \ell}(y) &:=& K_{\ell }(x-y)-K_{\delta \ell }(x-y)\;.
        \label{hLcutoffdef}
\eeqn
We remark that $\hat K_\ell$ is a product of differences of
Fej\'er kernels, and that
for $x\in\LL$ and $\delta>0$, $R_{x,\delta, \ell}(y)$
is an approximate  characteristic function supported on a
cubical shell of side length $2\ell$ centered at $x$, and thickness
$(1-\delta)\ell$.

The author thanks H.-T. Yau and
L. Erd\"os for the following observation, which
is the key to linking the localization length of eigenvectors to the
dynamics generated by $H_\omega$.
For a fixed realization of the random potential,
let $\{\psi_\alpha^{(L)}\}$ denote an orthonormal basis in
$\ell^2(\LL)$
of eigenfunctions of $H_\omega$ restricted to $\LL$,
\eqn
        (H_\omega-e_\alpha^{(L)})\psi_\alpha^{(L)}&=&0 \; {\rm on } \; \LL
        \;\;\;{\rm and}
        \nonumber\\
        \psi_\alpha^{(L)}&=&0 \; {\rm on} \;\partial\LL
        :=\Lambda_{L+1}\setminus\LL\;,
        \label{eigenL}
\eeqn
for
\eqn
        \alpha\in\alg_L&:=&\{1,\dots,|\LL|\}
        \nonumber\\
        e_\alpha^{(L)}&\in&\R \;.
\eeqn
For $\e$ small, let
\eqn
        \Dell:=\big\{\alpha\;\big|\;
        \sum_{x\in\LL} |\psi_\alpha^{(L)}(x)| \;
        \big\| R_{x, \delta, \ell} \psi_\alpha^{(L)}
        \big\|_{\ell^2(\LL)} < \e\;\big\} \;\subset\;\alg_L\;.
\eeqn
Then, $\{\psi_\alpha^{(L)}\}_{\alpha\in\Dell}$ contains the class of
exponentially localized eigenstates concentrated in
balls of radius $O(\frac{ \delta \ell }{ \log \ell })$ or smaller,
where we emphasize that $\delta$ is independent of $\ell$.
The additional factor $\log \ell$ in the denominator
compensates a volume factor $O(\ell^{3/2})$,
which arises due to the fact that $|\psi_\alpha^{(L)}(x)|$
appears only linearly, and not quadratically in the sum.
Our main result is the following theorem.

\begin{theorem}\label{mainthm}
Assume $L\gg\lambda^{-2}$, and that
$\{\psi_\alpha^{(L)}\}$ is an orthonormal $H_\omega$-eigenbasis in
$\ell^2(\LL)$,
satisfying (~\ref{eigenL})
with $\alpha\in\alg_L$, and $e_\alpha^{(L)}\in\R$.
Then, for $\lambda^{\frac{14}{15}}<\delta<1$ and
$\e_\delta:=\delta^{\frac37}$,
$$
        \Exp\lb
        \frac {|\alg_L\setminus\alg_{L,\e_\delta,\delta,\lambda^{-2}}^{(\omega)}|}
        {|\alg_L|}\rb\ge 1 - C \delta^{\frac{3}{14}}
        -C \lambda^{-2}L^{-1} 
        \;,
$$
for finite constants $C$ that are uniform in $L,\delta,\lambda$.
Furthermore,
$$
            {\Bbb P}\lb \liminf_{L\rightarrow\infty}
        \frac{|\alg_L\setminus
        \alg_{L,\e_\delta,\delta,\lambda^{-2}}^{(\omega)}|}{|\alg_L|}
        \geq 1- C \delta^{\frac{3}{14}}
        \rb = 1\;
$$
for $\lambda>0$  sufficiently small, and a finite constant $C$
that is uniform in $\lambda$ and $\delta$.
\end{theorem}

We note that in contrast to the results for dimension $d=1,2$
established in \cite{shscwo}, there is no restriction in dimension 3
on the range of values of $e_\alpha^{(L)}$. Furthermore, we
emphasize that the
correction to the lower bound of order $O(\lambda^{-2})$ on the
localization length is only logarithmic, while
the bound obtained in \cite{shscwo} for $d=2$ is of
order $O(\lambda^{-2+\e})$, for any arbitrary $\e>0$.

\section{Proof of the main theorem}
\label{sectionIII}

Key to Theorem {~\ref{mainthm}} is
the following lemma, which establishes a link between the localization length
of eigenvectors of $H_\omega$ and the dynamics generated by $H_\omega$.

\begin{lemma}\label{ceylemma}
(Joint with L. Erd\"os and H.-T. Yau)
Let $\{\psi_\alpha^{(L)}\}$ denote an orthonormal basis in
$\ell^2(\LL)$,
consisting of eigenvectors of $H_\omega$ satisfying (~\ref{eigenL}),
and assume that $1\ll\ell\ll L$.
Suppose that there exists $t>0$, such that for all $x\in\Z^3$,
\eqn\label{mainest}
            \Exp \lb\big\| R_{x, \delta, \ell}  e^{-i t H_\omega }
        \delta_x\big\|_{\ell^2(\Z^3)} ^2\rb \ge 1- \e  \;
\eeqn
is satisfied for some $\e=\e(\delta,\ell,t)>0$. Then,
$$
        \Exp\lb
        \frac {|\cDell| } {|\alg_L|}\rb\ge 1 - 2
        \e^{\frac12}-C \ell L^{-1}\;,
$$
for a constant $C$ which is independent of $\ell,L,\e$.
\end{lemma}

\prf
We have
\eqnn
        \delta_x &=& \sum_\alpha {a_x^\alpha} \psi_\alpha^{(L)}
        \; \;\\
        a_x^\alpha &=& \overline{\big\langle \delta_\vx \, , \,
        \psi_\alpha^{(L)} \big\rangle }
        = \overline{\psi_\alpha^{(L)}(x)} \;,
\eeqnn
so that in particular,
\eqn
        \|\delta_x\|_{\ell^2(\LL)}^2=\sum_{\alpha\in\alg_L}|a_x^\alpha|^2=1\;.
        \label{axalphl2norm}
\eeqn
By the Schwarz inequality, we get
\eqn
        \Big\| R_{x, \delta, \ell} e^{-i t H_\omega }
        \delta_x\Big\|_{\ell^2(\LL)}^2 
        &\leq& (1+ \frac1\eta )
        \Big\| R_{x, \delta, \ell} e^{-i t H_\omega }
        \sum_{\alpha \in \Dell}
        {a_x^\alpha} \psi_\alpha^{(L)} \Big\|_{\ell^2(\LL)}^2
        \nonumber\\
        &+&
        (1+\eta) \;
        \Big\| R_{x, \delta, \ell} e^{-i t H_\omega }
        \sum_{\alpha \in \cDell}
        {a_x^\alpha} \psi_\alpha^{(L)} \Big\|_{\ell^2(\LL)}^2 \; .
\label{CSest1}
\eeqn
For the first term on the r.h.s., we find
\eqn
        \Big\|R_{x, \delta, \ell}e^{-i t H_\omega }
        \sum_{\alpha \in \Dell}{a_x^\alpha}
        \psi_\alpha^{(L)} \Big\|_{\ell^2(\LL)}^2 
        &\leq& \Big\|R_{x, \delta, \ell}
        \sum_{\alpha \in \Dell}
        e^{-i t e_\alpha^{(L)} }{a_x^\alpha}
        \psi_\alpha^{(L)}\Big\|_{\ell^2(\LL)}
        \nonumber\\
        &\leq&  \sum_{\alpha\in \Dell}
        |\psi_\alpha^{(L)}(x)| \big\|  R_{x, \delta, \ell}
        \psi_\alpha^{(L)} \big\|_{\ell^2(\LL)} \;,
\eeqn
using the a priori bound
\eqn
        \Big\|R_{x, \delta, \ell}e^{-i t H_\omega }
        \sum_{\alpha \in \Dell}{a_x^\alpha}
        \psi_\alpha^{(L)} \Big\|_{\ell^2(\LL)}^2 
        &\leq&
        \Big\|
        \sum_{\alpha \in \Dell}e^{-i t e_\alpha^{(L)} }{a_x^\alpha}
        \psi_\alpha^{(L)} \Big\|_{\ell^2(\LL)}^2
        \nonumber\\
        &=&\sum_{\alpha\in\Dell}|a_x^\alpha|^2\leq 1
        \;,
\eeqn
which follows from $\|R_{x, \delta, \ell}\|_\infty=1$,
orthonormality of
$\{\psi_\alpha^{(L)}\}_{\alpha\in\alg_L}$ on $\ell^2(\LL)$,
and (~\ref{axalphl2norm}).
For the second term on the r.h.s. of (~\ref{CSest1}),
we likewise find
\eqn
        \Big\| R_{x, \delta, \ell} e^{-i t H_\omega }
        \sum_{\alpha \in \cDell}
        {a_x^\alpha} \psi_\alpha^{(L)}
        \Big\|_{\ell^2(\LL)}^2 
        &\leq&
        \Big\|\sum_{\alpha \in \cDell}e^{-i t e_\alpha^{(L)} }
        {a_x^\alpha} \psi_\alpha^{(L)} \Big\|_{\ell^2(\LL)}^2
        \nonumber\\
        &=&
        \sum_{\alpha \in \cDell}|a_x^\alpha|^2
        \nonumber\\
        &=&
        \sum_{\alpha \in \cDell}
        |\psi_\alpha^{(L)}(x)|^2 \; .
\eeqn
Averaging over $x\in\LL$, we have
\eqn
        &&\LLinv \sum_{x\in\LL} \big\| R_{x, \delta, \ell}
        e^{-i t H_\omega }
        \delta_x\big\|_{\ell^2(\LL)}^2\nonumber\\
        &\leq& (1+\eta)\frac{1}{|\LL|}
        \sum_{\alpha \in \cDell} \sum_{x\in\LL}
        |\psi_\alpha^{(L)}(x)|^2
        \nonumber\\
        &+& (1+ \frac1\eta )\frac{1}{|\LL|} \sum_{\alpha\in \Dell}
        \sum_{x\in\LL} |\psi_\alpha^{(L)}(x)|
        \big\| R_{x, \delta, \ell} \psi_\alpha^{(L)}
        \big\|_{\ell^2(\LL)}
        \nonumber\\
        &\leq& (1+\eta)
        \frac {1}{|\LL|}\big|\cDell\big|
        \label{DbarDsplitest}\\
        &+&(1+\frac1\eta )\frac{1}{|\LL|} \sum_{\alpha\in \Dell}
        \sum_{x\in\LL} |\psi_\alpha^{(L)}(x)|
        \big\|R_{x, \delta, \ell} \psi_\alpha^{(L)}
        \big\|_{\ell^2(\LL)}\; .
        \nonumber
\eeqn
Let
\eqn
        S_{2\ell,L}:=\big\{x\,\big|\,\inf_{y\in\partial\LL}
        |x-y|\leq 2\ell  \big\}
\eeqn
and $\tLL:=\LL\setminus S_{2\ell,L}$, such that
$R_{x,\delta,\ell}\cap\partial\LL=\emptyset$ $\forall x\in \tLL$. Then,
\eqn
        &&\LLinv \sum_{x\in\LL}  \big\| R_{x, \delta, \ell}
        e^{-i t H_\omega } \delta_x\big\|_{\ell^2(\LL)}^2
        \nonumber\\
        &=&
        \frac{1}{|\tLL|} \sum_{x\in\tLL}  \big\| R_{x, \delta, \ell}
        e^{-i t H_\omega }
        \delta_x\big\|_{\ell^2(\Z^3)}^2+O( \ell L^{-1})\;,
\eeqn
by compactness of the support of
$R_{x, \delta, \ell}$.
By definition of $\Dell$, the last term in (~\ref{DbarDsplitest}) is
bounded by $(1+\frac1\eta)\e$. Thus, recalling that $|\LL|=|\alg_L|$,
\eqn
        \frac {|\cDell | } {|\alg_L|}
        &\geq& \frac{1}{1+\eta}
        \frac{1}{|\tLL|}\sum_{x\in\tLL}  \big\| R_{x, \delta, \ell}
        e^{-i t H_\omega } \delta_x\big\|_{\ell^2(\Z^3)}^2
        \nonumber\\
        &-& \frac{1+\frac1\eta}{1+\eta}\, \e  - c\ell L^{-1}\;.
        \label{fracAcAlowbd}
\eeqn
Taking expectations, using  (~\ref{mainest}),
and choosing $\eta=\e^{\frac12}$, the claim follows.
\endprf

\begin{lemma}\label{mainproblemma}
Under the same assumptions as in Lemma {~\ref{ceylemma}},
\eqn
        {\Bbb P} \lb\liminf_{L\rightarrow\infty}
        \frac {|\cDell| } {|\alg_L|}\ge 1 - 2 \e^{\frac12}\rb
        =1 \;.
\eeqn
\end{lemma}

\prf
We consider the family of translation operators
$\tau_x:\omega_y\mapsto\omega_{x+y}$,
for $x\in\Z^3$, which acts ergodically on
the probability space $(\Omega,{\mathcal F},{\Bbb P})$,
\cite{cyfrkisi}.

Let $U_{\tau_x}$ denote the
unitary translation operator $(U_{\tau_x}\phi)(y)=\phi(x+y)$ on
$\ell^2(\Z^3)$.
Then, clearly,
\eqn
        U_{\tau_x}^*H_{ \omega}U_{\tau_x}
        =-\frac12\Delta+\lambda V_{\tau_{-x}\omega}
        =H_{\tau_{-x}\omega}
\eeqn
with $V_{\tau_x \omega} (y)=V_\omega(x+y)$, and
\eqn
        &&\frac{1}{|\tLL|} \sum_{x\in\tLL}  \big\| R_{x, \delta, \ell}
        e^{-i t H_\omega } \delta_x\big\|_{\ell^2(\Z^3)}^2
        \nonumber\\
        &=&
        \frac{1}{|\tLL|} \sum_{x\in\tLL}  \big\|
        (U_{\tau_x}^*R_{x, \delta, \ell}U_{\tau_x})
        (U_{\tau_x}^* e^{-i t H_\omega }U_{\tau_x})
        \delta_0\big\|_{\ell^2(\Z^3)}^2
        \nonumber\\
        &=&
        \frac{1}{|\tLL|} \sum_{x\in\tLL} \big\| R_{0, \delta, \ell}
        e^{-i t H_{\tau_{-x} \omega} }
        \delta_0\big\|_{\ell^2(\Z^3)}^2 \;,
\eeqn
by unitarity of $U_{\tau_x}$. By the Birkhoff-Khinchin ergodic
theorem, applied to the random variable
$X(\omega):=\| R_{0, \delta, \ell}
e^{-i t H_{\omega} } \delta_0\big\|_{\ell^2(\Z^3)}^2$, we obtain,
for fixed $\lambda$,
\eqn
        \label{ergodic}\\
        \liminf_{L\rightarrow\infty}\frac{1}{|\tLL|} \sum_{x\in\tLL}
        \big\| R_{0, \delta, \ell}
        e^{-i t H_{\tau_{-x} \omega} } \delta_0\big\|_{\ell^2(\Z^3)}^2=
        \Exp \lb\big\| R_{0, \delta, \ell}  e^{-i t H_\omega }
        \delta_0\big\|_{\ell^2(\Z^3)}^2\rb
        \nonumber
\eeqn
with probability one. We note here that
clearly, the left hand side of (~\ref{mainest}) is independent of
$x\in\Z^3$.
Therefore, (~\ref{mainest}), (~\ref{fracAcAlowbd}) and
(~\ref{ergodic}) imply
\eqn
        {\Bbb P}\Big[\liminf_{L\rightarrow\infty}\frac {|\cDell|} {|\alg_L|}
        \ge 1 - \frac{\eta}{1+\eta}-
        \frac{2+\frac1\eta}{1+\eta}\;\e \Big]=1 \;,
\eeqn
and choosing $\eta=\e^{\frac12}$, the claim
follows.
\endprf

From here on, we will write
$\|\,\cdot\,\|_2\equiv\|\,\cdot\,\|_{\ell^2(\Z^3)}$.

To conclude the proof of Theorem {~\ref{mainthm}}, we use the key
Lemma {~\ref{fundestlemma} } below, which provides the lower bound
\eqn\label{Mainfullpropest}
        \Exp\lb\big\| R_{x,\delta,  \lambda^{-2}}
        e^{-i t(\delta,\lambda) H_\omega }
        \delta_x\big\|_2 \rb
        \geq 1 - C \delta^{\frac37}
\eeqn
for the choice
$t(\delta,\lambda)=\delta^{\frac67}\lambda^{-2}$,
$\lambda^{\frac{14}{15}}<\delta<1$,  and a constant
$c$ that is independent of $x$,   $\lambda$ and $\delta$.
Thus, choosing $\e=\delta^{\frac37}$, (~\ref{Mainfullpropest})
immediately implies Theorem {~\ref{mainthm}}.

\begin{lemma}\label{fundestlemma}
Let
\eqn
        t(\delta,\lambda)=\delta^{\frac67}\lambda^{-2} \;,
\eeqn
and $H_0:=-\frac12\Lap $.
Then, for  $\lambda$ sufficiently small,   $0<\delta<1$,
and all $x\in\Z^3$, the free evolution term satisfies
\eqn\label{fundest0}
        \big\|  R_{x,\delta,  \lambda^{-2}}
        e^{-i t(\delta,\lambda) H_0 } \delta_x \big\|_2
        \geq 1 -   C  \delta^{\frac37}   \;,
\eeqn
while
\eqn\label{fundest10}
        \Exp \Big[ \big\|   R_{x, \delta, \lambda^{-2}}
        \big[ e^{-i t(\delta,\lambda) H }-e^{-i t(\delta,\lambda) H_0 }
        \big]\delta_x  \big\|_2^2   \Big] 
        \leq
        C'  \delta^{\frac67} +  
        \delta^{-\frac{6}{49}}\lambda^{\frac27} \;,
\eeqn
for finite positive constants $C,C'$ that are independent of
$x$, $\lambda$ and $\delta$.
\end{lemma}

\prf
We may  assume that $x=0$.  Let
\eqn
        \ell_2:=\ell\;\;,\;\;\ell_1:=\delta\ell \;.
\eeqn
We recall that $(R_{0, \delta, \ell})^2= K_{\ell_2}^2-2K_{\ell_1}+K_{\ell_1}^2$.
To bound $\|K_{\ell_2}e^{-itH_0}\delta_0\|_2$, we note that
\eqn
        &&\Big|e^{-it\en(p)}-
        \int_{\Tor^3}dk \hat K_{\ell_2}(p-k) e^{-it\en(k)} \Big| 
        \nonumber\\
        &\leq& C\sup_{|p-k|\leq\gamma}\Big|e^{-it\en(p)}-
        e^{-it\en(k)}\Big|
        \nonumber\\
        &&+2\int_{\Tor^3} dk\; |\hat K_{\ell_2}(k)|
        \;\chi(|p-k|\geq\gamma)
        \nonumber\\
        &\leq& C\gamma t +C\gamma^{-1}\ell_2^{-1}
        \;,
        \label{fundest0K2err}
\eeqn
owing to 
\eqn
        |\hat K_{\ell_2}(k)|\leq C\prod_{j=1}^3
        \frac{\ell_2}{1+\|k_j\|_{\Z}^2\ell_2^2} \;\;,
        \;\;
        \int_{\Tor^3}dk \hat K_{\ell_2}(k)=1 \;,
\eeqn
where $\|r\|_{\Z}:=$dist$(r,\Z)$ for $r\in\R$,
which are basic properties of the Fej\'er kernel.
Thus, with
\eqn
        t=\delta^{\alpha}\lambda^{-2}
        \;\;,\;\;
        \ell_2=\lambda^{-2}
        \;\;,\;\;
        \gamma= t^{-\frac12}\ell_2^{-\frac12} \;,
\eeqn
we find
\eqn
        (~\ref{fundest0K2err})\leq C\delta^{\frac\alpha2} \;.
\eeqn
Hence,
\eqn
        \Big\| K_{\ell_2} e^{-itH_0}\delta_0\Big\|_2^2
        &=&\int_{\Tor^3}dp\;
        \Big|e^{-it\en(p)}+O(\delta^{\frac\alpha2}) \Big|^2
        \nonumber\\
        &\geq& 1-C\delta^{\frac\alpha2}  \;.
\eeqn
Next, we consider
\eqn
        \Big\| |K_{\ell_1}|^{\frac a2} e^{-itH_0}\delta_0\Big\|_2^2=
        \sum_{y\in \Z^3} \big|K_{\ell_1}(y)\big|^a
        \,\Big|\big(e^{-itH_0}\delta_0\big)(y)\Big|^2 \;,
        \label{fundest0K1bound}
\eeqn
where $a=1,2$, and
\eqn
        \big(e^{-itH_0}\delta_0\big)(y)=
        \int_{\Tor^3}dk e^{-i(t\en(k)-2\pi   k y)}\;.
        \label{Free-evol-phase-1}
\eeqn
We observe that the kinetic energy
$\en:\Tor^3\rightarrow[0,6]$ is
a real analytic Morse function with eight critical points in
the corners of the subcube $[0,\frac{1}{2}]^3\subset\Tor^3=[0,1]^3$.
Each of the remaining
critical points in $[0,1]^3\setminus[0,\frac{1}{2}]^3$ is identified
with one of the latter by symmetry.
The Hessians are diagonal and have entries
of modulus $4\pi^2$.

We bound $|(~\ref{Free-evol-phase-1})|$ by a stationary phase estimate.
For $|y|\leq C\ell_1$,
$\ell_1=\delta\lambda^{-2}$ and $t=\delta^\alpha\lambda^{-2}$,
it is clear that 
\eqn
		\nabla ( \en - 2\pi t^{-1} \langle y,\,\cdot\,\rangle)(k^*)=0
		\label{Phase-grad-1}
\eeqn
implies
$$
		|\nabla \en(k^*) |<C\delta^{1-\alpha}\;.
$$  
It follows that for each of the eight critical
points of $\en$, there is precisely one $k^*$ satisfying (~\ref{Phase-grad-1}) 
in its $\delta^{1-\alpha}$-vicinity, given that
$\delta^{1-\alpha}$ is sufficiently small.
Correspondingly, Hess$[\en](k^*)$ is in each of these cases
non-degenerate, with eigenvalues of modulus $O(1)$.

We introduce a smooth partition of unity $\sum\phi_j=1$ on $[0,1]^3$,
$j\in\{1,\dots,8\}$, continued over the boundary  
by periodicity, in a manner that each supp$\phi_j$ is centered at one
critical point of $\en$. 
By the above, a stationary phase estimate yields
$$
        \sup_{y\in{\rm supp}K_{\ell_1}}\Big|\sum_{j}
        \int_{\Tor^3} dk\; \phi_j(k)
        e^{-i(t\en(k)-2\pi  k y)}\Big| \leq C t^{-3/2}   \;.
$$ 
Consequently,
\eqn
        |(~\ref{fundest0K1bound})|\leq  C\ell_1^3 t^{-3}
        = C\delta^{3(1-\alpha)} \;,
\eeqn
and optimizing the bounds, we find $\alpha=\frac67$.

Our strategy to prove (~\ref{fundest10}) employs a modification of
the methods of L. Erd\"os and H.-T. Yau from \cite{erdyau}.
Thereby, we invoke a Duhamel expansion with remainder term,
and control the expectation by
classifying all contraction types occurring in the products of the
random potential. The remainder term is bounded by exploiting
the rarity of the event that a large number of
collisions occurs in a small time interval.

As a result, we obtain
\eqn\label{fundest1}
        \Exp \Big[ \big\| R_{x,\delta, \lambda^{-2}}\big[
        e^{-i t H }-e^{-i t H_0 } \big]\delta_x\big\|_2^2 \Big]
        \leq C_1\lambda^2 t +t^{-\frac{1}{7}} \;,
\eeqn
for a constant $C_1$ that is independent of
$x$, $\lambda$ and $\delta$.
This implies (~\ref{fundest10}) for the
asserted choice of $t$.
The proof of  (~\ref{fundest1}) will occupy sections
{~\ref{ranpotsubsec}} $\sim$ {~\ref{mainlemmaprf}}.
\endprf

\section{Expectation of products of random potentials}
\label{ranpotsubsec}

We shall to begin with consider the expectation of products of random potentials.
The pair correlation is given by the Kronecker delta
$$
        \Exp\big[\omega_{\vx_1}\omega_{\vx_2}\big]=
        \delta_{\vx_1,\vx_2}\;,
$$
and we recall that by our assumptions on $\omega_x$,
the $m$-point correlation is zero for any odd $m$.
The fourth order correlation yields
\eqn
        &&\Exp\big[\omega(x_1)\omega(x_2)\omega(x_3)\omega(x_4)\big]
        \nonumber\\
        &=&
        (1-\delta_{x_1, x_3})\delta_{x_1,x_2}\delta_{x_3,x_4}
        +(1-\delta_{x_1, x_2})\delta_{x_1,x_3}\delta_{x_2,x_4}\nonumber\\
        &&+
        (1-\delta_{x_1, x_3})\delta_{x_1,x_4}\delta_{x_2,x_3}
        +\tilde c_4\delta_{x_1,x_2}\delta_{x_3,x_4}\delta_{x_1, x_3}
        \label{exp4nonwick}\\
        &=& \delta_{x_1,x_2}\delta_{x_3,x_4}
        +\delta_{x_1,x_3}\delta_{x_2,x_4} +
        \delta_{x_1,x_4}\delta_{x_2,x_3}\nonumber\\
        &&
        +(\tilde c_4-3)\delta_{x_1,x_2}
        \delta_{x_3,x_4}\delta_{x_1, x_3}\;.\label{exp4wick}
\eeqn
The operation applied in passing from (~\ref{exp4nonwick}) to (~\ref{exp4wick})
will be referred to as {\em Wick ordering}.
By a renormalization of the fourth order moment of $\omega_x$,
$$
        \tilde c_4 \rightarrow  c_4 := \tilde c_4-3 \tilde c_2^2
$$
(where $\tilde c_2=1$), it decomposes (~\ref{exp4nonwick})
into independent terms.

For the Fourier transformed random potentials
$\hat\omega(k):=\sum_x\omega_x e^{2\pi i k x}$,
one obtains exact Dirac delta
distributions for the Wick ordered expression (~\ref{exp4wick}),
\eqnn
        &&\Exp\big[\hat\omega(k_1)\hat\omega(k_2)
        \hat\omega(k_3)\hat\omega(k_4)\big]
        \nonumber\\
        &=&\delta (k_1+k_2)\delta (k_3+k_4)
        +
        \delta (k_1+k_3)\delta (k_2+k_4)\\
        &&
        +\delta (k_1+k_4)\delta (k_2+k_3)
        + c_4\delta (k_1+k_2+k_3+k_4)\;.
\eeqnn
We note that this is not the case for the individual summands
in (~\ref{exp4nonwick}) prior to Wick ordering.
The same statement applies to all higher order correlations.

The Wick ordered product of an arbitrary
even number of random potentials is determined as follows.
We introduce, for $n,n'\in\N$ with $\bar n:=\frac{n+n'}{2}\in\N$, the
set
$$
        \cV_{n,n'}:=\Big\{1,\dots,n,n+2,\dots,n+n'+1\Big\} \;.
$$
In our later discussion, $\cV_{n,n'}$
labels a linearly ordered set of $n+n'$ random potentials
that are, in frequency space, subdivided
into a group of $n'$ copies of $\hat V_\omega$,
and a group of $n$ copies of $\overline{\hat V_\omega}$ (the complex
conjugate).
The label $n+1$ excluded here is reserved for a distinguished point
that is not attributed to a random potential. We note again that the
case
$n+n'\in2\N_0+1$ is trivial since all odd moments of $V_\omega$ vanish.

\begin{definition}
For $\bar n=\frac{n+n'}{2}\in\N$, let
$$
        \Pi_{n,n'}:=\bigcup_{m=1}^{\bar n}
        \Big\{\{S_j\}_{j=1}^m\Big||S_j|\in2\N
        ;\cV_{n,n'}=\cup_{j=1}^m
        S_j; S_j\cap S_{j'}=\emptyset
        \hbox{ {\rm if} } j\neq j'\Big\}\Big/\Sym_m
$$
denote the set of partitions of $\cV_{n,n'}$ into disjoint subsets
$S_j$ (referred to as blocks) of size
$|S_j|\in2\N$, where $\Sym_m$ is the $m$-th symmetric group.
Two partitions $\pi=\{S_j\}_{j=1}^m$, $\pi'=\{S'_j\}_{j=1}^m$,
are equivalent, $\pi=\pi'$, if $\exists \sigma\in\Sym_m$ such that
$S_j=S'_{\sigma(j)}$ for all $j\in\{1,\dots,m\}$.
A partition $\pi\in\Pi_{n,n'}$ will also be referred to as a
contraction (corresponding to contractions among random potentials).
\end{definition}

The number of  $\pi\in\Pi_{n,n'}$ consisting of $m$ blocks is given by
\eqn
        &&B_{\bar n}(m)
        \nonumber\\
        &&:=\sharp\Big\{\{S_j\}_{j=1}^m\Big|\cup_{j=1}^m S_j=\cV_{n,n'}
         ;|S_j|\in2\N;
        S_i\cap S_j=\emptyset\,{\rm if}\,i\neq j\Big\}\Big/\Sym_m
        \nonumber\\
        &&=
        \sum_{r=1}^{\bar n}
        \sum_{1\leq j_1,\cdots,j_r\leq \bar n}
        \sum_{1\leq l_1<\cdots<l_r\leq \bar n}
        \delta_{m,|\underline{j}|}
        \delta_{\bar n,\langle\underline{j},\underline{l}\rangle}
        \nonumber\\
        &&\hspace{2cm}\times\,
        \frac{(2\bar n)!} {((2l_1)!)^{j_1}
        \cdots((2l_r)!)^{j_r} } \frac{1}{(j_1!)\cdots(j_r!)} \;,
        \label{Bmdef}
\eeqn
where $\underline{j}:=(j_1,\dots,j_r)$,
$|\underline{j}|:=\sum_{i=1}^r j_i$, and
$\langle\underline{j},\underline{l}\rangle:=\sum_{i=1}^r j_i l_i$ for
every $r$.
Here, $j_i$ is the number of blocks of size $2l_i$. The factor
$\frac{1}{j_i!}$
arises because the order is irrelevant,
according to which blocks of the same size are counted.
We note that the number of partitions into products of pair correlators
(that is, $r=1$, $j=\bar n$, $l=1$) is
$$
        B_{\bar n}(\bar n)=1\cdot3\cdots(2\bar n-1)<
        2^{\bar n}(\bar n!) \;.
$$
On the other hand, it is clear that
$$
        B_{\bar n}(m)<\sum_{0\leq s_1,\dots,s_m\leq 2\bar n}
        \delta_{2\bar n, \sum_{i=1}^m s_i}
        \frac{(2\bar n)!}{(s_1!)\cdots(s_m!)}
        =m^{2\bar n} \;,
$$
hence for non-pairing contractions, i.e. $m<\bar n$,
$$
        \sum_{m=1}^{\bar n-1}B_{\bar n}(m) < \bar n^{2\bar n+1} \;.
$$
This trivial estimate will suffice for our purposes.

For $S\subset\cV_{n,n'}$, with $|S|\in2\N$, we define
$$
        \delta(x_{S}) := \sum_{y\in\Z^3}\prod_{j\in S}
            \delta_{x_j,y}\;,
$$
where $x_S:=(x_j)_{j\in S}$.
Then,
\eqn\label{ExpnonWick}
        \Exp \Big[ \prod_{j\in\cV_{n,n'}} V_{\omega}(\vx_j) \Big]
        &=&\sum_{m=1}^{\bar n}
        \sum_{\stackrel{\pi\in\Pi_{n,n'}}
        {\pi=\{S_j\}_{j=1}^m}}
        \Big( \prod_{j=1}^m \tilde c_{|S_j|}\delta(x_{S_j})\Big)
        \nonumber\\
        &\times&
        \prod_{1\leq i<j\leq m}(1-\delta_{x_{\mu(i)},x_{\mu(j)}})
\eeqn
where for definiteness, $\mu(i):=\min\big\{q\Big|q\in S_i\big\}$
(clearly, one could choose any arbitrary element of $S_i$).
Due to the second product, the factors in
$\prod c_{|S_j|}\delta_{S_j}$ are not independent. We note that
\eqn
        \delta(x_{S_{i}}) \delta(x_{S_{j}})
        \delta_{x_{\mu(i)},x_{\mu(j)}}
        =\delta(x_{S_i\cup S_j})\;,
        \label{SiSjunion}
\eeqn
where of course, $|S_i\cup S_j|=|S_i|+|S_j|$.
Therefore, expanding $\prod(1-\delta_{x_{\mu(i)},x_{\mu(j)}})$
in (~\ref{ExpnonWick}), using (~\ref{SiSjunion}) recursively,
and collecting all terms belonging to the same blocks, we find
\eqn\label{Krondeltprod}
        \Exp \Big[ \prod_{j\in\cV_{n,n'}} V_{\omega}(\vx_j) \Big]
        = \sum_{m=1}^{\bar n}
        \sum_{\stackrel{\pi\in\Pi_{n,n'}}
        {\pi=\{S_j\}_{j=1}^m}}
        \prod_{j=1}^m c_{|S_j|}\delta(x_{S_j}) \; ,
\eeqn
where the cumulant formula
\eqnn
        c_{2k}&=&\sum_{m=1}^k\sum_{r=1}^{k}
        \sum_{1\leq j_1,\cdots,j_r\leq k}
        \sum_{1\leq l_1<\cdots<l_r\leq k}
        \delta_{m,|\underline{j}|}
        \delta_{k,\langle\underline{j},\underline{l}\rangle}
        \\
        &&\hspace{2cm}\times\,
        \frac{(-1)^{m-1}(2k)!} {((2l_1)!)^{j_1}
        \cdots((2l_r)!)^{j_r} }
        \frac{\tilde c_{2l_1}^{j_1}\cdots
        \tilde c_{2l_r}^{j_r}}{(j_1!)\cdots(j_r!)} \;
\eeqnn
determines the renormalized moments of $\omega_x$.
Thus, (~\ref{Krondeltprod}) decomposes the expectation value into the
sum of
all possible products of correlators, which are now mutually
independent.
We observe that by (~\ref{omcorrdef}),
\eqn\label{renormc2kbound}
        |c_{2k}|&\leq&(2k)!\;k\;\sum_{r=1}^k\Big(
        \sum_{1\leq j_1,\cdots,j_r\leq
        k}\frac{\com^{j_1+\dots+j_r}}
        {(j_1!)\cdots(j_r!)}\Big)\Big(
        \sum_{1\leq l_1<\cdots<l_r\leq k}1\Big)
        \nonumber\\
        &\leq&
        (2k)!\;2k\; \sum_{r=1}^k \frac{(ke^{ \com})^r}{r!}
        \nonumber\\
        &\leq&k^{k+1}\frac{(2k)!}{k!}\;e^{k e^{\com}}
        \nonumber\\
        &\leq&(2k \; e^{\frac12 e^{\com} })^{2k+1}\;.
\eeqn
Then, the expectation of the full product of random potentials
decomposes
into
\eqn\label{expprodvomega}
                \Exp \Big[ \prod_{j\in\cV_{n,n'}}
        \hat V_{\omega}(\vk_j) \Big]
                = \sum_{m=1}^{\bar n}
                \sum_{\stackrel{\pi\in\Pi_{n,n'}}
        {\pi=\{S_j\}_{j=1}^m}}
                \Big(\prod_{j=1}^m c_{|S_j|}\Big)
        \delta \Big(\sum_{i\in S_j}k_i\Big)\;
\eeqn
in momentum space, where $c_{|S_j|}$ are the
renormalized moments of $\omega_x$.

\section{Duhamel Expansion}

Our aim is to prove the bound (~\ref{fundest10}) by classifying
and estimating the integrals corresponding to
all contractions occurring on the left hand side of
(~\ref{fundest10}).

To this end,
we invoke the Duhamel expansion of $\phi_t=e^{-it H_\omega}\delta_x$.
For $N\in\N$ large, which remains to be determined, it is given by
\eqn
        \big( e^{-itH_\omega}\delta_x-e^{-itH_0}\delta_x\big)(y)&=&
        \Big((-i\lambda)\int_0^t ds  e^{-i (t-s) H_\omega }V_\omega
        e^{-i s H_0 } \delta_x\Big)(\vy)
        \nonumber    \\
        &=&\sum_{n=1}^N \phi_{n,t} (\vy) +\Rem_{N,t}(\vy) \; .
	\label{Duh-def-1}
\eeqn
Writing
$$
        \Big[ \prod_{j=0}^n ds_j   \Big]_t :=  ds_0\cdots ds_n\,
        \delta(\sum_{j=0}^n s_j-t)
$$
for brevity, the Fourier transform of the $n$-th Duhamel term is given by
\eqn
        \hat\phi_{n,t}(\vk_0)
        &=&  (-i\lambda)^n
        \int_{\R_+^{n+1}}
        \Big[ \prod_{j=0}^n ds_j   \Big]_t
        \int_{(\Tor^3)^n}   e^{2\pi i \vk_n\cdot\vx}
        e^{- i \sum_{j=0}^n s_j \en(\vk_j)}
        \nonumber\\
        &\times&
        \prod_{j=1}^n
        \hat V_\omega(\vk_j-\vk_{j-1})dk_j
        \nonumber\\
        &=& \frac{ie^{\e t}
        (-i\lambda)^n}{2\pi}   \int_\R d\alpha e^{- i \alpha t}
        \int_{(\Tor^3)^n}
        e^{2\pi i \vk_n\cdot\vx}
        \prod_{j=0}^n
        \frac{1}{\en(\vk_j)-\alpha-i\e}
        \nonumber\\
        &\times&\prod_{l=1}^n
        \hat V_\omega(\vk_l-\vk_{l-1}) dk_l\; ,
\eeqn
where we shall choose
\eqn
        \e=t^{-1}
\eeqn
in all that follows.
$\alpha$ is an energy parameter,
and the multiplication operators $\frac{1}{\en(k_j)-\alpha-i\e}$
are the Fourier transformed resolvents of $-\frac12\Lap$
(which will also be referred to as particle propagators).
The explicit formula for the remainder term $\Rem_{N,t}$ can be found
in (~\ref{remainderterm}) below.

We note that in this analysis, $\e=t^{-1}$ and $\lambda$ will be the small
parameters of the theory,
which will ultimately be related through $\e=C\lambda^2$.

Let $\Hpl_-:=\{z\in\C\big|\Im(z)\leq0\}$.
The integrand is analytic in $\alpha$, and it is not
hard to see that the path of the $\alpha$-integration
can, for any fixed $n\in\N$, be deformed
away from $\R$ into the closed contour
\eqn
        I=I_\R\cup I_{\Hpl_-} \label{defIloop}
\eeqn
with
\eqnn
        I_\R &:=& [-1, 7]\\
        I_{\Hpl_-}&:=& ([-1, 7]-i)\cup (-1-i(0,1]) \cup (7-i(0,1])
        \subset \Hpl_-\;,
\eeqnn
which encloses ${\rm spec}\big\{-\frac{1}{2}\Lap -i\e\big\} =
[0,6]-i\e$.
Consequently,
\eqnn
        \hat\phi_{n,t}(\vk_0)&=&
        \frac{ie^{\e t} (-i\lambda)^n}{2\pi} \int_I d\alpha e^{- i \alpha t}
        \int_{(\Tor^3)^n} dk_1\cdots dk_n\,
        e^{2\pi i \vk_n\cdot\vx}
        \\
        &\times&\Big[\prod_{j=0}^n
        \frac{1}{\en(\vk_j)-\alpha-i\e}\Big]
        \prod_{l=1}^n\hat V_\omega(\vk_l-\vk_{l-1}) \;,
\eeqnn
where the loop $I$ is taken in the clockwise direction.

Using the Schwarz inequality,
\eqn
        l.h.s. \; of \; (~\ref{fundest1})\leq
        2 \Exp\Big[ \big\| \sum_{n=1}^N \phi_{n,t} \big\|_2^2 \Big]
        +2\Exp\Big[ \big\| \Rem_{N,t}^\e \big\|_2^2 \Big] \; .
\eeqn
For $1\leq n,n' \leq N$, and $\bar n:=\frac{n+n'}{2}\in\N$, we have
\eqn
        \Exp\lb \langle\phi_{n',t},\phi_{n,t}\rangle \rb
        &=&  \frac{e^{2\e t} \lambda^{2n}}{(2\pi)^2}
        \int_{I\times \bar I} d\alpha d\beta e^{-it(\alpha-\beta)}
        \nonumber\\
        &\times&
        \int_{(\Tor^3)^{2\bar n+2}} \Big[\prod_{j=0}^n
        \prod_{l=0}^{n'} dk_j  d\tilde k_l\Big] \;
        \delta (k_0-\tk_0)e^{-2\pi i(\vk_n-\tilde\vk_{n'})\cdot\vx}
        \nonumber\\
        &\times&
        \prod_{j=0}^n\prod_{l=0}^{n'}
        \frac{1}{\en(k_j)-\alpha-i\e}
        \frac{1}{\en(\tk_l)-\beta+i\e}
        \nonumber\\
        &\times&
        \Exp\Big[\prod_{j=1}^n\prod_{l=1}^{n'} \hat V_\omega(k_j-k_{j-1})
        \overline{\hat V_\omega(\tk_l-\tk_{l-1})}\,\Big]
        \; ,
\eeqn
where $\bar I$ is the complex conjugate of $I$, and taken in the
counterclockwise direction by the variable $\beta$.

Introducing new variables
\eqn
        \up&=&(p_0,\dots,p_n,p_{n+1},\dots,p_{2\bar n+1})
        \\
        &:=&(\tk_{n'},\dots,\tk_0,\vk_0,\dots,\vk_n)
\eeqn
and
\eqn
        (\alpha_j,\sigma_j) &=& \left\{\begin{array}{ll}(\alpha,1)&
        0\leq j\leq n\\
        (\beta,-1)&n<j\leq 2n+1 \;, \end{array}\right.
\eeqn
we can write
\eqn
        \Exp[\langle \phi_{n',t},\phi_{n,t} \rangle ]
        &=& \frac{e^{2\e t} \lambda^{2\bar n}}{(2\pi)^2}
        \int_{I\times\bar I} d\alpha d\beta e^{-it(\alpha-\beta)}
        \nonumber\\
        &\times&
        \int_{(\Tor^3)^{2\bar n+2}} d\up \;\delta (p_n-p_{n+1})
        \prod_{j=0}^{2\bar n+1}\frac{1}{\en(p_j)-\alpha_j-i\sigma_j\e}
        \nonumber\\
        &\times&
        \Exp\Big[\prod_{\stackrel{i=1}{i\neq n+1}}^{2\bar n+1}
        \hat V_\omega(p_i-p_{i-1})\Big]
        e^{-2\pi i(p_0-p_{2\bar n+1})\cdot\vx}
        \; , \label{exppotgenexpr}
\eeqn
noting that $\overline{\hat V(k)}=\hat V(-k)$.

Let $\pi=\{S_j\}_{j=1}^m\in\Pi_{n,n'}$ denote a partition. Let
$$
        \delta_{S_j} (\up):=
        \delta \Big(\sum_{i\in S_j}(p_i - p_{i-1})\Big)
$$
and
$$
	\delta_\pi (\up):=\prod_{j=1}^m \delta_{S_j} (\up) \;.
$$
Then, the contribution to  (~\ref{exppotgenexpr}) corresponding to
$\pi$ is given by the singular integral
\eqn
        \amp& := &
        \frac{e^{2\e t} \lambda^{2\bar n}}{(2\pi)^2}
        \int_{I\times\bar I} d\alpha d\beta e^{-it(\alpha-\beta)}
        \int_{(\Tor^3)^{2\bar n+2}} d\up\;
        \delta (p_n-p_{n+1})
        \nonumber\\
        & \times&
        e^{-2\pi i (p_0-p_{2\bar n+1})\cdot\vx}
        \Big(\prod_{l=1}^m c_{|S_l|} \Big)
        \delta_\pi (\up)\nonumber\\
        &\times&
        \prod_{j=0}^{2\bar n+1} \frac{1}{\en(p_j)-\alpha_j-i\sigma_j\e}
        \;,
        \label{mainpairinteg}
\eeqn
referred to as the {\em (Feynman) amplitude} corresponding to $\pi$.
The expectation (~\ref{exppotgenexpr}) is obtained from
summing the amplitudes $\amp$ over all partitions $\pi\in\Pi_{n,n'}$.

\section{The graph representation of contractions}
\label{graphsection}

To estimate the expectation (~\ref{exppotgenexpr}), it is necessary
to classify the singular integrals $\amp$, whose size depends
on the structure of $\pi$.
For this combinatorial problem, it is natural to represent $\pi$,
encoded in the delta distributions $\delta_\pi$
in (~\ref{mainpairinteg}), by {\em (Feynman) graphs}.
We shall use the following prescription,
cf. Figure 1. We draw two parallel
solid 'particle lines', joined together at one end, accounting for
$\delta(p_n-p_{n+1})$, containing $n$, respectively $n'$
vertices, where $n+n'\in2\N$. Every pairing contraction is depicted by
a dashed
line joining the respective vertices. The higher correlation
contractions
corresponding to $\delta_{S_j}$ are represented
by $|S_j|\in2\N$ dashed lines connecting the corresponding vertices to
one mutual vertex that is disjoint from the particle lines.
Any (solid) edge that lies on a particle line refers to a
particle propagator.

Let $G_\pi$ denote the graph associated to a partition
$\pi=\{S_j\}_{j=1}^m\in \Pi_{n,n'}$.
The set of vertices of the graph $G_\pi$ is denoted by $V(G_\pi)$,
and the set of edges as $E(G_\pi)$. $V(G_\pi)$ is the union
$V(G_\pi)=V_p(G_\pi)\cup V_{hc}(G_\pi)$,
where $V_p(G_\pi)=\cV_{n,n'}$ is the $n+n'$-subset of vertices on the
particle line,
and $V_{hc}(G_\pi)$ is the subset of vertices disjoint from the
particle lines,
which are associated to correlations of higher order than two.
In the product of
Kronecker deltas (~\ref{Krondeltprod}) in the position space picture,
the elements of $V_p(G_\pi)$ correspond to the
sites $x_i$ of random potentials, while the elements of $V_{hc}(G_\pi)$
correspond to the dummy summation variables $y_j$.

\begin{definition}
A contraction $\pi=\{S_j\}\in\Pi_{n,n'}$ is called a pairing
contraction if $m=\bar n$, so that $|S_j|=2$ for all $j$.
Otherwise, $\pi$ is called a higher (order)
correlation contraction, or a type III contraction
(cf. Definition {~\ref{deltatypesdef}} below).
\end{definition}

It is in fact necessary to introduce the following finer classification
of families of contractions, see \cite{erdyau}.

\begin{definition}\label{deltatypesdef}
The delta distributions associated to partitions $\pi\in\Pi_{n,n'}$ of
the set $\cV_{n,n'}$ are classified into
the following types, according to the corresponding subgraph structure.
A delta distribution $\delta_{S}(\up)=\delta(p_{i}-p_{i-1}+p_{j}-p_{j-1})$
associated to a pairing $S$ with $|S|=2$ is of
\\
{\sc Type I}  if
$i,j\leq n$.
\\
{\sc Type I'} if
$i,j > n$.
\\
{\sc Type II} if
$i\leq n$, but $j\geq n+2$.
\\
A delta function $\delta_{S}$ is of
\\
{\sc Type III} if $|S|\geq4$, that is, if it is not associated to a
pairing contraction.
\end{definition}

Hence, a partition of $\cV_{n,n'}$ is of type III if it contains a
type III delta distribution.

\begin{definition}
A pairing contraction $\pi\in\Pi_{n,n'}$ is called crossing if
$\delta_\pi$
contains two delta distributions
$\delta(p_{i_1}-p_{i_1-1}+p_{j_1}-p_{j_1-1})$ and
$\delta(p_{i_2}-p_{i_2-1}+p_{j_2}-p_{j_2-1})$, with $j_r>i_r$, such
that
$i_1-i_2$ and $j_1-j_2$ have the same signs.
\end{definition}

\begin{definition}
A non-crossing pairing contraction $\pi\in\Pi_{n,n'}$ is called
nested if $\delta_\pi$ contains two delta distributions
$\delta(p_{i_1}-p_{i_1-1}+p_{j_1}-p_{j_1-1})$ and
$\delta(p_{i_2}-p_{i_2-1}+p_{j_2}-p_{j_2-1})$, with $j_r>i_r$,
both
either of type I or of type I', such that
$i_1-i_2$ and $j_1-j_2$ have opposite signs.
\end{definition}

\begin{definition} A non-crossing and non-nested pairing
contraction is called simple.
A simple pairing contraction is called a ladder graph if
all of its associated delta functions are of type II.
\end{definition}

Assume $\pi\in\Pi_{n,n'}$ is a pairing contraction.
A spanning tree $T$ of $G_\pi$ is a connected tree graph that contains
$V(G_\pi)$. We denote the set of edges contained in $T$ by $E_T$,
and refer to the corresponding momenta as tree momenta.
The momenta corresponding to the edges in the complement $E_L=E_T'$
are referred to as loop momenta. Adding any edge of $E_L$ to the
spanning tree $T$ produces a loop.

\begin{definition}
A spanning tree $T$ of $G_\pi$ with $\pi=\{S_j\}_{j=1}^m\in\Pi_{n,n'}$
a pairing contraction is
called complete if it contains all contraction lines,
and the edge corresponding to the momentum $p_n$,
but not the one corresponding to the momentum $p_{n+1}$.
\end{definition}

\section{Simple pairing contractions}

It is our aim to estimate $|\amp|$ for each type of contractions
$\pi\in\Pi_{n,n'}$ listed above.
We shall proceed by first discussing simple pairings, then crossing and
nested pairings, and finally type III contractions.

Similarly as in \cite{erdyau}, we will find that the amplitudes
$\{\amp\big|\pi\;{\rm simple}\}$ completely dominate over those
associated to all other contraction classes
(notably even in the presence of type III contractions).

\subsection{The ladder graph}

The simplest member in the class of simple pairings in
$\Pi_{n,n}$ is the {\em ladder graph}.
It corresponds to the pairing
$\pi=\{S_j\}_{j=1}^n\in\Pi_{n,n}$, with $S_{j}=\{j,2n+2-j\}$,
such that $|S_j|=2$, and
\eqn
        \amp&=&\frac{e^{2\e t} \lambda^{2n}}{(2\pi)^2}
        \int_{I\times\bar I}  d\alpha d\beta e^{-it(\alpha-\beta)}
        \int_{(\Tor^3)^{2n+2}} \dup\;
        \delta(p_n-p_{n+1})
        \nonumber\\
        &\times&
        e^{-2\pi i(p_0-p_{2n+1})\cdot\vx}
        \prod_{l=0}^{2n+1} \frac{1}{\en(p_l)-\alpha_l-i\sigma_l\e}
        \nonumber\\
        &\times&
        \prod_{j=1}^{n}
        \delta \Big((p_j-p_{j-1})+(p_{2n+2-j}-p_{2n+1-j})\Big)\;,
        \label{ladderCpidef}
\eeqn
cf. (~\ref{mainpairinteg}).

The following $L^\infty$ and $L^1$ resolvent estimates will be
used extensively in the sequel.

\begin{lemma}\label{propboundslemma}
Let $0<\e\ll1$. Then,
\eqn\label{resolvbound1}
        \sup_{\alpha\in I}\;\sup_{p\in\Tor^3}\;
        \frac{1}{|\en(p)-\alpha-i\e|}&\leq& \frac1\e
        \\
        \label{resolvbound2} \sup_{\alpha\in I}
        \int_{\Tor^3} dp \;  \frac{1}{|\en(p)-\alpha-i\e|}
        &\leq& C \log \frac1\e
        \nonumber\\
        \sup_{p\in\Tor^3} \int_I |d\alpha| \;
        \frac{1}{|\en(p)-\alpha-i\e|}
        &\leq& C \log \frac1\e
\eeqn
for finite constants $C$ that are uniform in $\e$.
\end{lemma}

\prf
Since by definition of $I$,
$\inf_{p\in\Tor^3}\dist(\en(p)-i\e,I)=\e $,
and since $|I|$ is finite, (~\ref{resolvbound1}) and the second
estimate
in (~\ref{resolvbound2}) are evident.

To prove the first estimate in (~\ref{resolvbound2}), we first show
that
the measure of the isoenergy surface
$$
        \Sigma_\alpha:= \{p\in \Tor^3\big|\;\en(p)=\alpha \}
$$
is uniformly bounded with respect to $\alpha\in I\cap\R$.
We note that for $\Im(\alpha)\neq0$, the asserted bound is trivial.
For $p=(p_1,p_2,p_3)\in\Tor^3$, let
\eqn
        \etwo (\up):=\sum_{j=1,2}(1-\cos2\pi p_j)
        \;,\;\;
        \up:=(p_1,p_2)
\eeqn
denote the Fourier transform of the 2-D nearest neighbor Laplacian,
and
\eqn
	s_r:=\{\up\in\Tor^2\big|\;\etwo (\up)=r \}
\eeqn
the corresponding level curves. Then, 
\eqn
        \mes \{s_r\}=\int_{[0,1]^2}d\up \, \delta(\etwo (\up)-r)
        \label{A2mes-1}
\eeqn
is easily seen to be uniformly bounded,
\eqn
        \sup_r\mes \{s_r\}<C\; .
        \label{A2mes-2}
\eeqn
Therefore,
\eqn
        \mes\{\Sigma_\alpha\}&=&\int_0^1 dp_3 \int_{\Tor^2} d\up
        \, \delta(\etwo (\up)+(1-\cos2\pi p_3)-\alpha)
	\nonumber\\
        &=&2\int_0^1 dk(1-k^2)^{-\frac12} \int_{\Tor^2}  d\up
        \,\delta(\etwo (\up)+1-k-\alpha)
	\nonumber\\
        &\leq&2\int_0^1 dk(1-k^2)^{-\frac12} \sup_r\mes \{s_r\}
        \;\leq\;C \;,
\eeqn
uniformly in $\alpha$.
Thus, defining
\eqn
        R_j(\alpha,\e):=
        \Big\{p\in\Tor^3\Big|
        2^{j}\e<\en(p)\leq 2^{j+1}\e\Big\} \;,
\eeqn
we have
\eqn
        \mes\{R_j(\alpha,\e)\}&=&
        \int_{2^j\e}^{2^{j+1}\e}d\alpha'\int_{\Tor^3}dp
        \delta(\alpha'-\en(p))
        \nonumber\\
        &=&\int_{2^j\e}^{2^{j+1}\e}d\alpha'\mes\{\Sigma_{\alpha'}\}
        \nonumber\\
        &\leq&2^{j}\;\e\;\sup_{\alpha'\in \R}\mes\{\Sigma_{\alpha'}\}
        \;\leq\;C 2^{j} \e
\eeqn
for a constant $C$ that is independent of $j$,
$\e$, and $\alpha$.
Hence, introducing a dyadic decomposition of $\Tor^3$ with respect
to $\en$ centered about $\Sigma_\alpha$, we find
\eqn
        \int_{\Tor^3}dp\; \frac{1}{|\en(p)-\alpha-i\e|}
        &\leq&\sum_{j}\int_{R_j(\alpha,\e)}dp\;
        \Big|\frac{1}{\en(p)-\alpha-i\e}\Big|
	\nonumber\\
        &\leq&C \sum_{j}
        \frac{\mes\{R_j(\alpha,\e)\}}{2^{j}\e}
        \nonumber\\
	&\leq&C\log\frac{1}{\e}\;,
\eeqn
for $0\leq j\leq  C\log\frac{1}{\e}$ and a constant $C$ that is uniform in
$\e$ and $\alpha$, as claimed.
\endprf

\begin{lemma}
Let
\eqnn
        K^{(n)}(\vp_0,\dots,\vp_n;t)
        &:=&\int ds_0\dots ds_n
        \delta\Big(t-\sum_{r=0}^n s_r\Big)
        e^{-i\sum_{j=0}^n s_j \en(\vp_j)}\\
        &=&\frac{ie^{\e t}}{2\pi}\int_I d\alpha e^{-i\alpha t}
        \prod_{j=0}^n\frac{1}{\en(\vp_j)-\alpha-i\e}\;.
\eeqnn
Then there exists a finite constant $C_\mu$ for every $0<\mu<1$ such that
\eqn
        \big\|K^{(n)}(\,\cdot\,;t)\big\|_{L^2((\Tor^3)^{n+1})}^2\leq
        \frac{(C_\mu t)^n}{(n!)^\mu}\;.
\eeqn
\end{lemma}

\prf
Clearly,
\eqnn
        \big\|K^{(n)}(\,\cdot\,;t)\big\|_{L^\infty((\Tor^3)^{n+1})} 
        &\leq&\int_{\R_+^{n+1}} ds_0\dots ds_n
        \delta(t-\sum s_r)
        =\frac{t^n}{n!} \;.
\eeqnn
Furthermore,
\eqn
        \big\|K^{(n)}(\,\cdot\,;t)\big\|_{L^{2-\mu}
        ((\Tor^3)^{n+1})}^{2-\mu} 
        &\leq&C \int_{(\Tor^3)^{n+2}} \dup\; \delta(p_n-p_{n+1})
        \Big[\int_I |d\alpha| \prod_{j=0}^n
        \frac{1}{|\en(\vp_j)-\alpha-i\e|}\Big]^{2-\mu}
        \nonumber\\
        &\leq& C \int_{(\Tor^3)^{n+2}}\dup\; \delta(p_n-p_{n+1})
        \int_I |d\alpha| \prod_{j=0}^n
        \frac{1}{|\en(\vp_j)-\alpha-i\e |^{2-\mu}}
        \nonumber\\
        &\leq& C_\mu^n \; t^{n(1-\mu)} \;,
\eeqn
for a finite constant $C_\mu$.
The claim then follows from interpolation.
\endprf

We conclude that the ladder contribution can be estimated by
\eqn
        &&\lambda^{2n}\int_{(\Tor^3)^{2n+2}}
        d\up \;
        \prod_{i=1}^{n}
        \delta (p_i -p_{2n+1-i})
        \nonumber\\
        &&\hspace{2cm}
        \times\,K^{(n)}(p_0,\dots,p_{n};t)
        \overline{ K^{(n)}( p_{n+1},\dots, p_{2n+1};t)}
        \nonumber\\
        &\leq&  \lambda^{2n}
        \big\|K^{(n)}(\,\cdots\,;t)\big\|_{L^2((\Tor^3)^{n+1})}^2
        \; \leq \;  \frac{(C_\mu \lambda^2 t)^n}{(n!)^\mu}
\eeqn
for  $0<\mu<1$. It is clear that the product of delta distributions
appearing here is equivalent to the one in (~\ref{ladderCpidef}).

\subsection{Immediate recollisions}

We next estimate general simple pairings which include all possible
combinations of type I and I' contractions. Given any type I or type I'
delta function $\delta(p_i-p_{i-1}+p_j-p_{j-1})$ in a simple
pairing graph, where $j>i$, one
necessarily finds $i=j-1$. Otherwise,
either a crossing or a nesting pairing occurs. Hence, any
type I or I' delta function in a simple pairing reduces to
$\delta(p_{i+1}-p_{i-1})$, for some $i$.

\begin{definition}
A type I or type I' pairing of the form $\delta(p_{i+1}-p_{i-1})$
is called  an immediate recollision.
\end{definition}

The subintegral in $\amp$ corresponding to an immediate recollision is
given by either
\eqn
        \Th(\alpha,\e):=
        \int_{\Tor^3} \frac{dq}{\en(q)-\alpha-i \e}
\eeqn
or $\Th(\beta,-\e)$.
It contributes to a renormalization of the particle propagator,
see \cite{erdyau}, and satisfies
the following estimates, which will be of extensive use.

\begin{lemma}\label{Thetalemma}
For $\alpha\in I$,
\eqn
        \sup_{\alpha\in I}\sup_{\e>0}|\Th(\alpha,\e) |&<&C \;
        \label{Thest-1}
\eeqn
and
\eqn
        \label{deralmThinf}
        |\partial_\alpha^m\Thinf(\alpha,\e)| &\leq&   C \;
        \e^{-(m-1/2)} \;(m!  )
        \\ 
        |\Th(\alpha,\e)-\Th(\alpha',\e)|&\leq& C  \e^{-1/2}
        |\alpha-\alpha'|
        \label{ThDiffest}
\eeqn
for finite constants $C$ that are independent of $m,\e,\alpha,\alpha'$,
and $m\in\N$.

\end{lemma}

\prf
We recall that $\alpha\in I= I_\R\cup I_{\Hpl_-}$ from
(~\ref{defIloop}).
The case $\alpha\in I_{\Hpl_-}$ is trivial.
For $\alpha\in I_\R=[-1,7]$, we write
$$
        \Thinf(\alpha,\e)=\int_{\R_+} ds \int_{\Tor^3} dp \,
        e^{-is(\en(p)-\alpha-i\e)}  \;,
$$
and recall that
$\en:\Tor^3\rightarrow[0,6]$ is a real analytic
Morse function with eight critical points.
We choose a smooth partition of unity $1=\sum\phi_j$ on
$\Tor^3$, $j\in\{1,\dots,8\}$, requiring that the support of
each $\phi_j$ is centered
about precisely one critical point of $\en$, so that
\eqn
        \Thinf(\alpha,\e)&=&\sum_{j}
        \int_{\R_+} ds \int_{\Tor^3} dp \,\phi_j(p) \,
        e^{-is(\en(p)-\alpha-i\e)}  \;.
\eeqn
Using a stationary phase estimate, we find
\eqn
        |\Thinf(\alpha,\e)|&<& \sum_{j} C_j \int_{\R^+}ds\,
        e^{-\e s} (1+s)^{-\frac32}   \;,
\eeqn
where the constants $C_j$ are independent of $\e$ and $\alpha$.
This proves (~\ref{Thest-1}).

Likewise,
\eqn
        \partial_\alpha^m\Thinf(\alpha,\e)&=& \sum_{j}
        \int_{\R_+} ds \int_{\Tor^3} dp  \phi_j(p)
        (is)^m e^{-is(\en(p)-\alpha-i\e)}\;.
\eeqn
Thus
\eqn
        |\partial_\alpha^m\Thinf(\alpha,\e)|&<&
        \sum_{j}C_j \int_{\R^+}ds\,s^m e^{-\e s}
        (1+s)^{-\frac32}  \;,
\eeqn
which implies (~\ref{deralmThinf}), and for $m=1$, also (~\ref{ThDiffest}).
\endprf

\subsection{General simple pairings}

In a more general context, simple pairings comprise
progressions of neighboring immediate recollisions
on each particle line before and after each type II contraction.
In this sense, simple pairings are ladder graphs that are
decorated with immediate recollisions on the propagator lines.
Let us assume that there are $q$ neighboring delta functions of type
I', starting
at the particle propagator carrying the momentum $p_i$. Then, $\amp$
contains the corresponding subintegral
\eqnn
        &&\int_{(\Tor^3)^{2q}}   dp_{i+1}\cdots dp_{i+2q}
        \Big(\prod_{l=i}^{i+2q}\frac{1}{\en(p_l)-\alpha-i\e} \Big)
        \prod_{j=1}^q \delta(p_{i+2j+1}-p_{i+2j-1})
        \nonumber\\
        &&\hspace{2cm}=
        \frac{\Th(\alpha,\e)^q}{(\en(p_i)-\alpha-i\e)^{q+1}}  \;.
\eeqnn
The analogous expression for a  progression of $q$ neighboring
delta functions of type I is obtained from substituting
$\alpha\rightarrow\beta$ and $\e\rightarrow-\e$.

Let us consider a simple pairing $\pi\in\Pi_{n,n'}$ which contains $m$
type II contractions. Let
\eqnn
        \qm &:=&(q_0,q_1,\dots,q_m) \in\N_0^{m+1}
\eeqnn
and
\eqnn
        |\qm |&:=& q_0+\cdots+q_m \;,
\eeqnn
and, for $n-n'\equiv0$ (mod 2),
\eqn
        A_{n,n'}:=\big\{m\in\N_0\big|m-n\equiv0\;
        ({\rm mod}\;2)\;,\;m\leq \min\{n,n'\}\big\} \;.
        \label{Annsetdef}
\eeqn
The sum over all simple pairings at fixed $n$
gives (after reindexing the momentum variables)
\eqn
        \sum_{\stackrel{\pi\in\Pi_{n, n'}}
        {\pi\;{\rm simple }}}\amp
        &=&\sum_{m\in A_{n,n'}}\frac{\lambda^{2m}e^{2\e t}}{(2\pi)^2}
        \int_{I\times \bar I} d\alpha d\beta e^{-it(\alpha-\beta)}
        \nonumber\\
        &\times&
        \sum_{|\qm |  =\frac{n-m}{2}}
        \sum_{|\tqm  |=\frac{n'-m}{2}}
        \int_{(\Tor^3)^{m+1}} dp_0\cdots dp_m
        \nonumber\\
        &\times&
        \prod_{i=0}^m\frac{(\lambda^2\Th(\alpha,\e))^{q_i}}
        {(\en(p_i)-\alpha-i\e)^{q_i+1}}
        \frac{(\lambda^2\Th(\beta,-\e))^{\tilde q_i}}
        {(\en(p_i)-\beta+i\e)^{\tilde q_i+1}}  \;.
        \label{simplepairfixedn}
\eeqn
Let us comment on this expression, cf. Figure 2.
For $i=1,\dots,m$, $p_{i-1}$ is the momentum preceding, and $p_i$
the momentum following the $i$-th type II pairing.
Notably, a direct recollision conserves the momentum.
For $1\leq i\leq m$, $q_i$ and $\tilde q_i$ are the
numbers of neighboring type I and I' pairings after the $i$-th type
II contraction.
$q_0$ and $\tilde q_0$ are the number of neighboring type I and I'
pairings before the $1$-st
type II pairing.

Clearly, all $n-m$ random potentials on each particle line not
involved in type II
contractions are part of type I, respectively type I' pairings
(immediate recollisions).
Since each immediate recollision contracts precisely two random
potentials, the sum
over $m$ takes steps of size 2, such that $m\in A_{n,n'}$.
Therefore,
\eqn
        |\qm |=\frac{n-m}{2}
        \;,\;\;
        |\tqm  |= \frac{n'-m}{2}
\eeqn
is clear.
In particular, $m=n=n'$ corresponds to the ladder graph.

\begin{lemma}\label{simplepairlemmleqN}
For fixed $n,n'$ with $\bar n:=\frac{n+n'}{2}\in\N$,
the contribution of the sum of all simple pairings is
bounded by
\eqn
        \Big|\sum_{\pi\in\Pi_{n,n'}\;{\rm simple}}\amp\Big| 
        &\leq&
        \delta_{n,n'}
        \frac{(C_0 \lambda^2  \e^{-1})^{ \bar n}}{(\bar n!)^{1/2}} +
        \bar n^2 \e^{\frac12} |\log \e|
        \big(C \e^{-1} \lambda^2 |\log \e|\big)^{\bar n}  \;,
\eeqn
and for $\lambda^2 \e^{-1}\leq1$,
\eqn
        \sum_{n,n'=1}^N \Big| \sum_{\pi\in\Pi_{n,n'}\;{\rm simple}}\amp\Big| 
        &\leq& C_1  \lambda^2 \e^{-1} + \e^{\frac12} N^3
        \big(C\lambda^2 \e^{-1} |\log \e|\big)^{N } \;,
\eeqn
where  $C,C_0,C_1$ are uniform in $N$ and $\e$, and where $C_0$ and
$C_1$
are defined in (~\ref{c0def}).
\end{lemma}

\prf
Let us assume for fixed $n,n'$ under the stated conditions that
$\pi\in\Pi_{n,n'}$ is simple, and contains $m$ type II pairings. Let
\eqn
        \amp = \ampl_{main}[\pi] + \ampl_{error}[\pi] \;,
\eeqn
where
\eqn
        \ampl_{main}[\pi] &:=&
        \frac{ e^{2\e t}\lambda^{2m}}{(2\pi)^2}
        \int_{I\times \bar I} d\alpha\; d\beta \;e^{-it(\alpha-\beta)}
        \nonumber\\
        &\times&
        \sum_{|\qm |  =\frac{n-m}{2}}
        \sum_{|\tqm  |=\frac{n'-m}{2}}
        \int_{(\Tor^3)^{m+1}}  dp_0\cdots dp_m
        \nonumber\\
        &\times&
        \prod_{i=0}^m
        \frac{(\lambda^2\Th(\en(p_0),\e))^{q_i}}
        {(\en(p_i)-\alpha-i\e)^{q_i+1}}
        \frac{(\lambda^2\Th(\en(p_0),-\e))^{\tilde q_i}}
        {(\en(p_i)-\beta+i\e)^{\tilde q_i+1}}  \;.
\eeqn
Then, recalling (~\ref{Annsetdef}),
\eqn
        \sum_{\pi\in\Pi_{n,n'}\;{\rm simple}}\ampl_{main}[\pi]
        &=&
        \sum_{m\in A_{n,n'}}
        \frac{e^{2\e t} \lambda^{2m}}{(2\pi)^2}
        \int_{I\times \bar I} d\alpha\; d\beta\; e^{-it(\alpha-\beta)}
        \label{eq43}\\
        &\times&
        \sum_{|\qm |  =\frac{n-m}{2}}
        \sum_{|\tqm  |=\frac{n'-m}{2}}
        \int_{(\Tor^3)^{m+1}}    dp_0\cdots dp_m
        \nonumber\\
        &\times&
        \prod_{i=0}^m\frac{(\lambda^2\Th(\en(p_0),\e))^{q_i}}
        {(\en(p_i)-\alpha-i\e)^{q_i+1}}
        \frac{(\lambda^2\Th(\en(p_0),-\e))^{\tilde q_i}}
        {(\en(p_i)-\beta+i\e)^{\tilde q_i+1}}   \;.
	\nonumber
\eeqn
Let $p_j^{(q_j)}=(p_j,\dots,p_j)$ ($q_j$ copies), and
$\dupm:=dp_0\cdots dp_m$.

We note that
\eqn
        \int_{(\Tor^3)^{m+1}}\dupm
            \Big|K^{(\frac{m+n}{2})}(p_0^{(q_0+1)},
        \dots, p_m^{(q_m+1)};t)\Big|^2 \leq
            \frac{(C_\mu t)^{\frac{m+n}{2}} }{(\frac{m+n}{2}!)^{\frac12}} \;.
\eeqn
Thus, by the Schwarz inequality,
\eqn
        &&\int_{(\Tor^3)^{m+1}} d\up^{(m+1)}
        K^{(\frac{n+m}{2})}(p_0^{q_0+1},\dots,p_m^{q_m+1};t)
        \nonumber\\
        &&\hspace{4cm}\times\;
        \overline{
        K^{(\frac{n'+m}{2})}(p_0^{\tilde q_0+1},\dots,
        p_m^{\tilde q_m+1};t)}
        \nonumber\\
        &\leq& \frac{(Ct)^{\frac{m+\bar n}{2}} }
        {(\frac{m+n}{2}!)^{\frac14}(\frac{m+n'}{2}!)^{\frac14}} \;.
\eeqn
Therefore,
\eqn
        &&\\
        |(~\ref{eq43})|&\leq&
        \sum_{m\in A_{n,n'}}\sum_{|\qm |= \frac{n-m}{2}}
        \sum_{|\tqm |= \frac{n'-m}{2}}
        \frac{2\lambda^{2m}(Ct)^{\frac{m+\bar n}{2}}
        \big|\lambda^2\Th(p_0;\e)\big|^{\bar n-m}}
        {(\frac{m+n}{2}!)^{\frac14}(\frac{m+n'}{2}!)^{\frac14}}
        \;.
        \nonumber
\eeqn
By
\eqn
        \sum_{ |\qm | = \frac{n-m}{2} }1 <C^{n-m }
	\label{Simplexmn2bound}
\eeqn
and Lemma {~\ref{Thetalemma}},
\eqn
        \Big|\sum_{\pi\in\Pi_{n,n}\;{\rm simple}} \ampl_{main}[\pi]\Big|
        &\leq&   \sum_{m\in A_{n,n'}}
        \frac{\lambda^{2m} (C\lambda^2)^{\bar n-m}(Ct)^{\frac{m+\bar n}{2} } }
        {(\frac{m+n}{2}!)^{\frac14}(\frac{m+n'}{2}!)^{\frac14} }
        \nonumber\\
        &\leq&
        \delta_{n,n'}
        \frac{(C\lambda^2  t)^{\bar n}}{(\bar n!)^{\frac12}}
        + C \bar n t^{-1}  (C\lambda^2 t)^{\bar n}    \;,
\eeqn
for finite constants $C$ that are independent of $\e$.
The first term after the second inequality sign accounts for
the ladder graph in $\Gamma_{n,n}$, corresponding to the case
$m=n=n'=\bar n$, as we recall.

Next, we consider the error term
\eqn
        &&\sum_{\pi\in\Pi_{n,n}\; {\rm simple}} \ampl_{error}[\pi]
        =\frac{1}{(2\pi)^2}\int_{I\times\bar I}
        d\alpha d\beta \; e^{-it(\alpha-\beta)}
        \nonumber\\
        &&\hspace{1cm} \times  \,
        \sum_{m\in A_{n,n'}}
            \sum_{|\qm |  =\frac{n-m}{2}}
        \sum_{|\tqm  |=\frac{n'-m}{2}}
        \lambda^{2m+2\sum_{j=0}^m(q_j+\tilde q_j)}
        \label{iidefeq}\\
        &&\hspace{1cm} \times\,
        \int_{(\Tor^3)^{m+1}}\dupm
        \prod_{i=0}^m
        \frac{1}{(\en(p_i)-\alpha-i\e)^{q_i+1}
        (\en(p_i)-\beta+i\e)^{\tilde q_i+1}}
        \nonumber\\
        &&\hspace{1cm}  \times\,
        \left[ \; \Th(\alpha,\e)^{\frac{n-m}{2}}\,
        \Th(\beta,-\e)^{\frac{n'-m}{2}}  -
        \Th(\en(p_0),\e)^{\frac{n-m}{2}}\,
        \Th(\en(p_0),-\e)^{\frac{n'-m}{2}} \;  \right] \;.
        \nonumber
\eeqn
Lemma {~\ref{Thetalemma}} implies that difference
in $[\cdots]$ on the last line is bounded by
$$
        \Big[\frac{n-m}{2}|\en(p_0)-\alpha|+
        \frac{n'-m}{2}|\en(p_0)-\beta|\Big]
        \e^{-1/2} C^{\bar n-m-1} \;,
$$
for a constant $C$ independent of $\e$.
Thus,  we arrive at
\eqnn
            |(~\ref{iidefeq})|&\leq&\bar n \; \e^{1/2}
        \sum_{m\in A_{n,n'}} (C\lambda^2  \e^{-1}|\log \e|)^{m}
        (C\lambda^2 \e^{-1}|\log \e|)^{\bar n-m}
        \\
        &\leq& \bar n^2 \, \e^{1/2}
        \Big(C \lambda^2  \e^{-1}|\log \e| \Big)^{\bar n}   \;.
\eeqnn
again using (~\ref{Simplexmn2bound}).

Summarizing, we have
\eqn
        \Big| \sum_{\pi\in\Pi_{n,n'}\;{\rm simple}} \amp\Big| &\leq&
        \delta_{n,n'}\frac{(C_0 \lambda^2  \e^{-1})^{\bar n}}{(\bar n!)^{\frac12}}
        +    n \e \big(C \lambda^2 \e^{-1}\big)^{\bar n }
        \nonumber\\
        &&\hspace{1cm}
        + \bar n^2 \e^{1/2}
            \big(C\lambda^2\e^{-1}|\log \e|\big)^{\bar n}   \;,
\eeqn
for some constant $C_0$. Furthermore, let
\eqn
        C_1:= \sum_{\bar n=1}^\infty \frac{C_0^{ \bar n}}
        {(\bar n!)^{\frac{1}{2}}}  \;.
        \label{c0def}
\eeqn
Then, for $\lambda^2\e^{-1}\leq1$,
$$
        \sum_{n=1}^N|(~\ref{simplepairfixedn})|\leq C_1
        \lambda^2 \e^{-1}
        +  \e^{1/2} N^3 \big( C \lambda^2 \e^{-1}
        |\log \e|\big)^{N }   \;,
$$
where the constants $C_0,C_1,C$ are uniform in $N$, $\lambda$, $\e$.
\endprf

We remark that for general $T:=\lambda^2t=\lambda^2\e^{-1}>1$,
the constant $C_1$ in the above estimate would be replaced by
$C^T$, where $C$ is uniform in $\lambda$ and $\e=t^{-1}$.

\subsection{A priori bound on pairing graphs}

All pairing graphs obey the
following a priori bound.

\begin{lemma}
Let $\pi\in\Pi_{n,n'}$ be a pairing graph, and
$\bar n:=\frac{n+n'}{2}\in\N$. Then,
\eqn
        |\amp|\leq |\log \e|^3 (C \lambda^2\e^{-1}|\log \e|)^{\bar n} \;.
        \label{aprioribdpair}
\eeqn
\end{lemma}

\prf
For the detailed argument, we refer the reader to \cite{erdyau}.
One chooses a complete spanning tree $T$ on $\pi$, and
estimates the propagators supported on $T$ in $L^\infty$.
Using
the bounds in Lemma {~\ref{propboundslemma}}, one obtains
a factor $\e^{-\bar n}$. The
loop propagators are estimated in $L^1$,
and yield a factor $(C|\log\e|)^{\bar n+3}$.
\endprf

\subsection{Crossing and nested pairings}

We shall next prove that for all $\pi\in\Pi_{n,n}$ which contain a crossing
or nested pairing contraction, $|\amp|$ is
a factor $O(\e^{\gex })$
smaller than the a priori bound (~\ref{aprioribdpair}) on pairing graphs.
This is sufficient to
compensate the factor $n!$ accounting for the number of pairing contractions.

\begin{lemma}
The sum of all crossing and nested pairing contractions in
$\Pi_{n,n'}$ (where $\bar n:=\frac{n+n'}{2}\in\N$) is bounded by
$$
        \sum_{\pi\in\Pi_{n,n'}\;{\rm crossing\;or\;nested}} |\amp|\leq
        \bar n! \, \e^{\gex } |\log \e|^3  (C \lambda^2\e^{-1}|\log \e|)^{\bar n} \; .
$$
\end{lemma}

\prf
By lemmata {~\ref{crosslessN}} and {~\ref{nestlessN}} below, every
pairing
contraction of crossing or nesting type can be bounded by
$$
                 (C \lambda^2 \e^{-1} |\log \e|)^{\bar n} \e^{\gex } |\log \e|^3 \;,
$$
and clearly, there are at most $2^{\bar n} \bar n!$ such graphs.
\endprf

\begin{lemma}\label{crosslessN}
Suppose that $\pi\in\Pi_{n,n'}$ corresponds to a pairing contraction that
contains at least one crossing, and that $\bar n=\frac{n+n'}{2}\in\N$. Then,
$$
        |\amp|\leq \e^{\gex }(C \lambda^2\e^{-1}|\log \e|)^{\bar n}  |\log \e|^3 \;.
$$
\end{lemma}

\prf
Let $T$ denote a complete spanning tree for the graph $G_\pi$, and
$T^c$ its complement.
As demonstrated in \cite{erdyau},
all momenta supported on $T$ can be expressed as linear combinations
of loop momenta
supported on $T^c$.
If there exists a crossing pairing, it is shown in \cite{erdyau} that
there is a tree momentum $p_r$ in $T$ that depends on at least
two loop momenta $p_j,p_l$ in $T^c$,
\eqn
        p_r=\pm p_j \pm p_l \pm w
    \label{crossmomrel-1}
\eeqn
where $w\in\Tor^3$ is a linear combination of momenta not depending
on $p_j,p_l$.
Writing $p\equiv p_j$, $q\equiv p_l$, and integrating out all
delta distributions determined by $\pi$ against momenta supported on $T$,
the amplitude $\amp$ can be written in the form
\eqn
        \amp&=&\frac{e^{2\e t}\lambda^{2\bar n}}{(2\pi)^2}
        \int_{I\times\bar I}d\alpha d\beta e^{-it(\alpha-\beta)}
        \\
        &\times&
        \int_{\Tor^{3|T^c|}}
        \;dp\;dq\;\Big[\prod_{p_j\in T^c\atop p_j\neq p,q}dp_j\Big]\;
        F_\pi(p_j\in T^c;\alpha,\beta;\e)
        \nonumber\\
        &\times&\frac{1}{(\en(p)-\alpha_1\pm i \e)\,(\en(q)-\alpha_2\pm i \e)
        (\en(p\pm q+w)-\alpha_3\pm i\e)}\;,
	\nonumber
\eeqn
where $\alpha_i\in\{\alpha,\beta\}$, for $i=1,2,3$, and $|T^c|$ is
the number of edges of $T^c$. $F_\pi$ contains all resolvents except
the three on the last line, which carry the moments singled out in
(~\ref{crossmomrel-1}). Using an $L^1-L^\infty$ bound with
respect to the variables $p$, $q$ and $\alpha$, $\beta$, we have
\eqn
        |\amp|&\leq& \lambda^{2\bar n}
        \sup_{\alpha\in I}\sup_{\beta\in\bar I}
        \sup_{w\in \Tor^3}A_\e(w;\alpha,\beta)
        \\
        &\times&\sup_{p,q}
        \int_{I\times\bar I}d\alpha d\beta \Big|
        \int_{\Tor^{3(|T^c|-2)}}
        \Big[\prod_{p_j\in T^c\atop p_j\neq p,q}dp_j\Big]
        F_\pi(p_j\in T^c;\alpha,\beta;\e)\Big|\;,
	\nonumber
\eeqn
where
\eqn
        \label{Aeps-def-1}
            A_\e(w;\alpha,\beta) 
	         &:=&\int_{(\Tor^3)^2}\frac{dp\,dq}
            {|\en(p)-\alpha_1\pm i \e|\,|\en(q)-\alpha_2\pm i \e|
            |\en(p\pm q+w)-\alpha_3\pm i\e|}\;.
        \nonumber
\eeqn
It is clear that $\alpha_i=\alpha_j$ for at least one pair of indices
$i\neq j$.  

Using the trivial bound $A_\e(w,\alpha,\beta)\leq c
\e^{-1}(\log\frac1\e)^2$, one obtains
\eqn
        |\amp|\leq   |\log \e|^3  (C \lambda^2\e^{-1}|\log \e|)^{\bar n} \;,
\eeqn
which is the a priori bound (~\ref{aprioribdpair})
on all pairing graphs. It is insufficient
because the number of crossing graphs is $O(\bar n!)$, and
$\bar n!|\log \e|^3  (C \lambda^2\e^{-1}|\log \e|)^{\bar n}$ is not summable in $\bar n$.
Gaining
an extra factor $\e^{\gex }$ will (in combination with our
treatment of the error term of the truncated Duhamel expansion)
allow us to compensate the large combinatorial factor $\bar n!$.

Exploiting the crossing structure of $\pi$,
Lemma {~\ref{Aeestlemma}} below provides the bound
\eqn
        \sup_{w\in\Tor^3}\,\sup_{\alpha\in I}\,
        \sup_{\beta\in\bar I}\,
        A_\e(w,\alpha,\beta) < C\,\e^{-\gexc }(\log\frac1\e)^2 \;,
\eeqn
which is a factor $\e^{\gex}$ smaller than the a priori estimate.

For the remaining part of $\amp$, excluding the propagators
corresponding to the indices $n$ and $n+1$,
$L^\infty$-bounds on propagators in $T$, and $L^1$-bounds on
propagators in $T^c$,
produce a factor $(C \lambda^2\e^{-1}|\log\e|)^{\bar n-1}$.
The propagators corresponding to the indices $n$ and $n+1$ contribute
a factor $(C\log \e^{-1})^2$, as in (~\ref{pnpn1estimate}) below.
A detailed exposition is given in \cite{erd,erdyau}.
\endprf

\begin{lemma}
\label{Aeestlemma}
Let $A_\e(w,\alpha,\beta)$ be defined as in
(~\ref{Aeps-def-1}). Then,
\eqn
        \sup_{w\in\Tor^3}\,\sup_{\alpha\in I}\,
        \sup_{\beta\in\bar I}\,
        A_\e(w,\alpha,\beta) < C \e^{-\gexc }(\log\frac1\e)^2 \;.
\eeqn
\end{lemma}

\prf
To bound  (~\ref{Aeps-def-1}), it is necessary to estimate the measure
of the intersection between tubular neighborhoods of 
level surfaces of the kinetic energy function
$\en$ where the singularities of the resolvents
in (~\ref{Aeps-def-1}) are concentrated. Because the level surfaces of
$\en$ are non-convex for the 3-dimensional lattice model,
this is a much more difficult task than in the continuum case, 
where the latter are spheres.
After completing this work, we learned that a similar but somewhat stronger 
estimate (with an exponent -3/4 instead of -4/5) was proven independently in 
\cite{erdsalmyau}. 

We shall interpret the 3-dimensional integral (~\ref{Aeps-def-1}) as an average
over 2-dimensional crossing integrals. 
Let
\eqn
        \up:=(p_1,p_2)&,&\uq:=(q_1,q_2) \; \; \in \;[0,1]^2
	\nonumber\\
        \etwo (\up)&:=&\sum_{j=1}^2\cos2\pi p_j
	\nonumber\\
        \alpha_j(k)&:=&-\cos2\pi k +3-\left\{
        \begin{array}{ll}
        \beta&{\rm if}\;j=1\\
        \alpha&{\rm if}\;j=2,3\;,
        \end{array}\right.
\eeqn
so that
\eqn
		A_\e(w;\alpha,\beta)
		&=&\int_{[0,1]}dp_3\int_{[0,1]}dq_3\int_{[0,1]^2}d\uq
		\frac{1}{|\etwo (\uq-\uw)-\alpha_1(q_3-w_3)+i\e|}
		\nonumber\\
		&\times&\int_{[0,1]^2}d\up 
		\frac{1}{|\etwo (\up)-\alpha_2(p_3)-i\e|\,
		|\etwo (\up-\uq)-\alpha_3(p_3-q_3)-i\e|}\;.
		\nonumber
\eeqn
The level curves 
\eqn
		s_\alpha:=\big\{\up\in[0,1]^2\big|\,\etwo (\up)=\alpha\big\} 
\eeqn
of the 2-dimensional
kinetic energy function $\etwo$ are convex, but there is
one exceptional value of the energy $\alpha=0$ for which the corresponding level curve  
$s_{\alpha=0}$ is the union of four line segments of zero curvature. 
The lack of curvature poses a well-known difficulty in 2 dimensional lattice models. 
In 3 dimensions, this problem is resolved through the average with respect to $p_3,q_3$ 
(relative to which small curvature is an event of small probability).

Let 
\eqn
		U_{\tau}&:=&\big\{(p_3,q_3) \big|\,\alpha_1(q_3),\alpha_2(p_3),
		\alpha_3(p_3-q_3)\in(-\tau,\tau)\big\}
		\nonumber\\
		U_{\tau}^c&:=&[0,1]^2\setminus I_\tau\;,
\eeqn
where $0<\tau\ll1$ remains to be optimized.
Then, clearly,
\eqn
		\mes\{U_\tau\}<C\sqrt\tau\;.
		\label{mes-Utau-est-1}
\eeqn
Correspondingly, let
\eqn
		A_\e(w;\alpha,\beta)=(A)+(B)
\eeqn
where
\eqn
		(A)&:=&\int_{U_\tau}dp_3 dq_3\int_{[0,1]^2}d\uq
		\frac{1}{|\etwo (\uq-\uw)-\alpha_1(q_3-w_3)+i\e|}
		\\
		&\times&\int_{[0,1]^2}d\up
		\frac{1}{|\etwo (\up)-\alpha_2(p_3)-i\e|\,|\etwo (\up-\uq)-\alpha_3(p_3-q_3)-i\e|}\;,
		\nonumber
\eeqn
and
\eqn
		(B)&:=& \int_{U_{\tau}^c}dp_3 dq_3\int_{[0,1]^2}d\uq
		\frac{1}{|\etwo (\uq-\uw)-\alpha_1(q_3-w_3)+i\e|}
		\\
		&\times&\int_{[0,1]^2}d\up
		\frac{1}{|\etwo (\up)-\alpha_2(p_3)-i\e|\,|\etwo (\up-\uq)-\alpha_3(p_3-q_3)-i\e|}\;.
		\nonumber
\eeqn 
Therefore, with (~\ref{mes-Utau-est-1}),
\eqn
		(B)&<& C\sqrt{\tau}  \sup_{\alpha_j}\int_{[0,1]^2}d\uq
		\frac{1}{|\etwo (\uq-\uw)-\alpha_1 +i\e|}
		\nonumber\\
		&&\times\int_{[0,1]^2}d\up
		\frac{1}{|\etwo (\up)-\alpha_2 -i\e|\,|\etwo (\up-\uq)-\alpha_3 -i\e|}
		\nonumber\\
		&<&\frac{C\sqrt{\tau}}{\e}(\log\frac1\e)^2\;.
\eeqn
Next, we decompose $(A)$ into $(A)=(A_1)+(A_2)$ with
\eqn
		(A_1)&:=&\int_{U_\tau}dp_3 dq_3\int_{[0,1]^2}d\uq\;
		\frac{\chi(|\etwo (\uq-\uw)-\alpha_1(q_3-w_3)|<\eta)}{|\etwo (\uq-\uw)-\alpha_1(q_3-w_3)+i\e|}
		\nonumber\\
		&\times&\int_{[0,1]^2}d\up
		\frac{\chi(|\etwo (\up)-\alpha_2(p_3)|<\eta)}
		{|\etwo (\up)-\alpha_2(p_3)-i\e|}
		\\
		&&\hspace{3cm}\times
		\frac{\chi(|\etwo (\up-\uq)-\alpha_3(p_3-q_3)|<\eta)}
		{|\etwo (\up-\uq)-\alpha_3(p_3-q_3)-i\e|}\;,
		\nonumber
\eeqn
where $0<\eta\ll1$ remains to be determined. Then clearly, the term complementary to $(A_1)$
satisfies
\eqn
		(A_2)<C\eta^{-1}(\log\frac1\e)^2\;,
\eeqn
because the integrand of $(A_2)$ contains at least one characteristic function of the
form $\chi(|\etwo (\uv)-\alpha_j(v_3)|>\eta)$, where $v=(\uv,v_3)$ denotes either $q-w$,
$p$, or $p-q$. The corresponding resolvent can be estimated by $\eta^{-1}$, while the
remaining two resolvents in $(A_2)$ can be bounded in $L^1$ by $c(\log\frac1\e)^2$.
 
To bound $(A_1)$, we note that
\eqn
		(A_1)&<&\Big(\frac\eta\e\Big)^{3}\eta^{-3}
		\int_{U_\tau}dp_3 dq_3\int_{[0,1]^2}d\uq\;
		\chi(|\etwo (\uq-\uw)-\alpha_1(q_3-w_3)|<\eta) 
		\\
		&\times&\int_{[0,1]^2}d\up
		\chi(|\etwo (\up)-\alpha_2(p_3)|<\eta)\,\chi(|\etwo (\up-\uq)-\alpha_3(p_3-q_3)|<\eta)
		\;.\nonumber
\eeqn
let $h_\eta$ denote a smooth bump function supported in a
$\eta$-vicinity of the origin in $[0,1]^2$ (periodically continued over
the boundaries), with 
\eqn
		\|h_\eta\|_{L^1([0,1]^2)}<10
\eeqn
and
\eqn
		|\Fou^{-1}(h_\eta)(\ux)|\left\{
		\begin{array}{ll}
		\approx C&|\ux|<\eta^{-1}\\
		<C|\ux|^{-1}&|\ux|\geq\eta^{-1}\;.
		\end{array}
		\right.
		\label{heta-def-1}
\eeqn
$\Fou$ denotes the Fourier transform, and $\ux\in\Z^2$ is the variable conjugate
to $\up$, respectively $\uq$.
Furthermore, let
\eqn
		\delta_{s_\alpha}^{(\eta)}(\up)&:=&h_\eta*\delta_{s_\alpha}(\up)\;,
\eeqn
where $\delta_{s_\alpha}(\up):=\delta(\etwo (\up)-\alpha)$.

Choosing $h_\eta$ appropriately, 
\eqn
		\frac{1}{\eta}\chi(|\etwo (\up)-\alpha|<\eta)<\delta_{s_\alpha}^{(\eta)}(\up) \;.
\eeqn
Thus, letting $T_{\uw} f(\uq):=f(\uq-\uw)$,
\eqn
		(A_1)&\leq&\Big(\frac\eta\e\Big)^{3}\sup_{|\alpha_j|>\tau}\sup_{\uw}\int_{[0,1]^2}d\uq \;
		\delta^{(\eta)}_{s_{\alpha_1}}(\uq-\uw)
		\int_{[0,1]^2}d\up\;
		\delta^{(\eta)}_{s_{\alpha_2}}(\up)\;
		\delta^{(\eta)}_{s_{\alpha_3}}(\up-\uq)
		\nonumber\\
		&=&\Big(\frac\eta\e\Big)^{3}\sup_{|\alpha_j|>\tau}\sup_{\uw}\big\langle
		T_{\uw}\delta^{(\eta)}_{s_{\alpha_1}} \;,\;
		\delta^{(\eta)}_{s_{\alpha_2}}\;*\;
		\delta^{(\eta)}_{s_{\alpha_3}}\big\rangle_{L^2([0,1]^2)}
		\nonumber\\
		&=&\Big(\frac\eta\e\Big)^{3}\sup_{|\alpha_j|>\tau}\sup_{\uw}\big\langle
		\Fou^{-1}(T_{\uw}\delta^{(\eta)}_{s_{\alpha_1}}) \;,\;
		\Fou^{-1}(\delta^{(\eta)}_{s_{\alpha_2}})\;
		\Fou^{-1}(\delta^{(\eta)}_{s_{\alpha_3}})\big\rangle_{\ell^2(\Z^2)}
		\nonumber\\
		&=&\Big(\frac\eta\e\Big)^{3}\sup_{|\alpha_j|>\tau}\sup_{\uw}\sum_{\ux\in\Z^2} 
		\Fou^{-1}(T_{\uw}\delta^{(\eta)}_{s_{\alpha_1}})(\ux)\;
		\Fou^{-1}(\delta^{(\eta)}_{s_{\alpha_2}})(\ux)\;
		\Fou^{-1}(\delta^{(\eta)}_{s_{\alpha_3}})(\ux)
		\nonumber\\
		&=&\Big(\frac\eta\e\Big)^{3}\sup_{|\alpha_j|>\tau}\sup_{\uw}\sum_{\ux\in\Z^2} 
		\big(\,\Fou^{-1}(h_\eta)(\ux)\,\big)^3
		\Fou^{-1}(T_{\uw}\delta_{s_{\alpha_1}})(\ux)\;
		\nonumber\\
		&&\hspace{3cm}\times\;
		\Fou^{-1}(\delta_{s_{\alpha_2}})(\ux)\;
		\Fou^{-1}(\delta_{s_{\alpha_3}})(\ux)\;,
\eeqn
by the Plancherel identity.
Next, we observe that if $|\alpha|>\tau$, the curvature of 
$s_\alpha\subset[0,1]^2$ is uniformly
bounded below by $C\tau$,  where the
constant $C$ is independent of $\tau$.
We thus have the curvature induced decay estimate
\eqn
		|\Fou^{-1}(\delta_{s_{\alpha}})(\ux)|<C (\tau|\ux|)^{-\frac12}\;,
		\label{rest-est-1}
\eeqn
which appears also in the context of restriction estimates in harmonic analysis, \cite{st}.
To arrive at (~\ref{rest-est-1}), one introduces a smooth partition of unity 
$1=\sum_{j=1}^N g_j$ on $s_\alpha$, which splits it into $N$ arcs $s_{\alpha,j}$, 
$j=1,\dots,N$, with, say, $N=10$. For fixed $j$, one introduces a local orthogonal
coordinate system $(v_1,v_2)$ where the origin lies on $s_{\alpha,j}$ 
(say at its center) with the $v_1$-axis tangent to $s_{\alpha,j}$.
Then, $s_{\alpha,j}$ is the graph of a smooth function $\phi_{\alpha,j}(v_1)$ with 
$\phi_{\alpha,j}(0)=0$, $\partial_{v_1}\phi_{\alpha,j}(v_1)=0$
and $|\partial_{v_1}^2\phi_{\alpha,j}(v_1)|>C\tau$. Let $\un:=\frac{\ux}{|\ux|}=(n_1,n_2)$,
where $\ux\in\Z^2$, and let $\up_j$ denote the location of the origin of the 
$\uv$-coordinate system with respect to the $\up$-coordinates. Then,
\eqn
		\Fou^{-1}(\delta_{s_{\alpha,j}}g_j)(\ux) 
		&&=
		e^{2\pi i \ux\up_j}\int dv_1 \, \big(1+(\partial_{v_1}\phi_{\alpha,j}(v_1))^2\big)^{\frac12} \,
		\tilde g_j(v_1) \,
		e^{2\pi i |\ux| \Phi_{\alpha,j}(\un,v_1)}\;,
		\nonumber
\eeqn
where $\tilde g_j(v_1):=g_j(v_1,\phi_{\alpha,j}(v_1))$, and
\eqn
		\Phi_{\alpha,j}(\un,v_1):= n_1v_1+n_2\phi_{\alpha,j}(v_1)\;.
\eeqn
First of all, if $\ux$ is parallel to the $v_2$-axis so that $n_1=0$, one has
$|\partial_{v_1}\Phi_{\alpha,j}(\un,0)|=0$ and 
\eqn
		|\partial_{v_1}^2\Phi_{\alpha,j}(\un,0)|>C\tau \;.
		\label{Phi-Hess-bd-1}
\eeqn
Hence by a stationary phase estimate,
\eqn
		|\Fou^{-1}(\delta_{s_{\alpha,j}}g_j)(\ux)|<C(\tau|\ux|)^{-\frac12}\;.
		\label{Fou-sj-est-1}
\eeqn 
If $\ux$ is close to being parallel to the $v_2$-axis, so that $|n_1|<C$
is sufficiently small (independently of $\tau$), one can find $v_1=v_1(\un)$, 
such that $\partial_{v_1}\Phi_{\alpha,j}(\un,v_1(\un))=0$. This follows from
an application of the implicit function theorem and (~\ref{Phi-Hess-bd-1}). 
Moreover, by the assumption on the curvature of $s_\alpha$, we have
$|\partial_{v_1}^2\Phi_{\alpha,j}(\un,v_1(\un))|>C\tau$.
Therefore, (~\ref{Fou-sj-est-1}) is valid for all $\ux$ such that $|n_1|<C$ is
sufficiently small. 
If $|n_1|>C$,
\eqn
		|\partial_{v_1}\Phi_{\alpha,j}(\un,v_1)|>C'>0\;,
\eeqn
since $|\partial_{v_1}\phi_{\alpha,j}(v_1)|=O(|v_1|)$.
This implies an even stronger decay bound than (~\ref{Fou-sj-est-1}),
by standard oscillatory integral estimates. 
Hence, we arrive at (~\ref{rest-est-1}).

Noting that 
\eqn
		|\Fou^{-1}(T_{\uw}\delta_{s_{\alpha}})(\ux)|=
		|\Fou^{-1}(\delta_{s_{\alpha}})(\ux)|\;,
\eeqn
we obtain
\eqn
		(A_1)&<&C\Big(\frac\eta\e\Big)^{3}
		\sum_{\ux\in\Z^2}\big(\,\Fou^{-1}(h_\eta)(\ux)\,\big)^3
		(\tau|\ux|)^{-\frac32}
		\nonumber\\
		&<&C\Big(\frac\eta\e\Big)^{3}\tau^{-\frac32}\eta^{\frac12}\;,
\eeqn
due to (~\ref{heta-def-1}).

We thus arrive at 
\eqn
		A_\e(w;\alpha,\beta)<C\Big(\frac\eta\e\Big)^{3}\tau^{-\frac32}\eta^{\frac12}
		+C\Big(\eta^{-1}+\frac{\sqrt\tau}{\e}\Big)(\log\frac1\e)^2\;.
\eeqn
Setting $\eta=\e^{\frac45}$, $\tau=\e^{\frac25}$, the
claim follows.
\endprf

\begin{lemma}\label{nestlessN}
Let $\pi\in\Pi_{n,n'}$ with $\bar n=\frac{n+n'}{2}\in\N$, correspond to a non-crossing pairing contraction
that
contains at least one nested subgraph. Then,
$$
        |\amp|\leq \e^{\frac12} (C\lambda^2\e^{-1}|\log\e|)^{\bar n}
$$
\end{lemma}

\prf
In this case, $\pi$ comprises a nested subgraph of length $1<q\leq
n-2$, and
$\amp$ thus contains a subintegral
\eqn
        N_q(\alpha,\e)\delta(p_{j+2q}-p_{j})
        &=&\int_{(\Tor^3)^{2q}}   dp_{j} \cdots dp_{j+2q-1}
        \prod_{l=j+1}^{j+2q-2}\frac{1}{\en(p_l)-\alpha-i\e}
        \nonumber\\
        &\times&\prod_{k=1}^{q-1}
        \lambda^2\delta(p_{j+2k+2}-p_{j+2k})
        \nonumber\\
        &\times&
        \lambda^2\delta\Big(p_{j+1}-p_{j}+p_{j+2q }-p_{j+2q-1}\Big)
            \;,
        \label{Nqdelta}
\eeqn
where
\eqn
        N_q(\alpha,\e) :=\lambda^2\int_{\Tor^3}dp\,
        \frac{(\lambda^2\Th(\alpha,\e))^{q-1} }
        {(\en(p)-\alpha-i\e)^q}  \;.
\eeqn
Since its interior does not contain further nested subgraphs,
we refer to it as a simple nest,  cf. Figure 3.
We note that $p_0,p_n,p_{n+1}$, and $p_{2\bar n+1}$ can
never appear in the interior of a nested subgraph. It is clear that
\eqn
        \big|N_q(\alpha,\e)\big|
        &\leq&\lambda^2 |\lambda^2\Th(\alpha,\e) |^{q-1}
        \Big|\frac{1}{(q-1)!}\partial_\alpha^{q-1}\int_{\Tor^3}
        dp\,\frac{1}{\en(p)-\alpha-i\e}\Big|
        \nonumber\\
        &\leq&
        \frac{\lambda^2 ( c  \lambda^2 )^{q-1}}{(q-1)!}
        \int_{\R_+}ds\,s^{q-1}e^{-\e s}(1+s)^{-\frac32}
        \nonumber\\
        &\leq&
        ( C \lambda^2\e^{-1})^{q}\e^{\frac32}   \;,
        \label{Nqalphaeest}
\eeqn
where we have used (~\ref{deralmThinf}).

Without any loss of generality, let us assume that the
simple nest has length $q$ (i.e. it contains $q$ immediate
recollisions),
and that the momentum with largest label preceding it is $p_j$, with
$j+2q<n$,
so that the expression corresponding to (~\ref{Nqdelta}) is
$N_q\delta(p_{j+2q}-p_j)$.

The contributions to $\amp$ stemming from the pairing contractions
outside of the simple nest can be estimated in the following way.
There are $2(\bar n-q)$ momenta not contained in the nest,
apart from those carrying indices in $J:=\{n, n+1, j+2q\}$.
Let $\pi'$ denote the graph obtained from $\pi$ by
removing the simple nest together with the edges labelled by $J$.
Let $T$ denote a spanning tree of $\pi'$ containing all of the
contraction lines,
and
$\bar n-q$ of the particle lines in $\pi'$.
The pairings supported on $\pi'$ can be written in the form
\eqnn
        &&\int_{(\Tor^3)^{n-q}}\Big[\prod_{r\in T^c}
        \frac{dp_r   }
        {(\en(p_r)-\alpha_r-i\sigma_r\e)^{\mu_j(r)}}\Big]
        \prod_{s\in T}\frac{1}
        {(\en(w_s)-\alpha_s-i\sigma_s\e)^{\mu_j(s)}} \;,
\eeqnn
where each $w_s\in\Tor^3$ is
a linear combination of momenta $p_{j_s}$ with $j_s\in T^c$.
Here, we have introduced
$\mu_j(r):=1+\delta_{j,r}$ to accommodate the fact that
the edge labelled by $j+2q$ is excluded from $\pi'$.
We correct this omission by using $p_{j+2q}=p_j$, which
is enforced by a delta distribution, and by squaring the propagator
corresponding to the edge with index $j$.

It then follows that 
\eqn
        |\amp|&\leq&\Big(\sup_{\alpha\in I}|N_q(\alpha,\e)|\Big)\lambda^{2(n-q)}
        \int_{I\times \bar I}|d\alpha|\,|d\beta|
        \nonumber\\
        &&\times\,
        \int_{\Tor^3}\;dp_n\;
        \frac{1}{|\en(p_n)-\alpha-i\e|^{\mu_j(n)}
        |\en(p_{n})-\beta+i\e| }
        \\
        &&\times\,
        \int_{(\Tor^3)^{\bar n-q}}\Big[\prod_{r\in T^c}\;dp_r\;
        \frac{1 }
        {|\en(p_r)-\alpha_r-i\sigma_r\e|^{\mu_j(r)}}\Big]
	\nonumber\\
	&&\hspace{3cm}\times
        \prod_{s\in T}
        \frac{1}{|\en(w_s)-\alpha_s-i\sigma_s\e|^{\mu_j(s)}} \;.
        \nonumber
\eeqn
Since
\eqn
        &&\int_{I\times \bar I}|d\alpha|\,|d\beta|
        \int_{\Tor^3}\;dp_n\;\frac{1}{|\en(p_n)-\alpha-i\e|^{\mu_j(n)}
        |\en(p_{n})-\beta+i\e| }
        \label{pnpn1estimate}\\
        &&\leq\; C \e^{1-\mu_j(n) }|\log\e|^2 ,\;\;
        \nonumber
\eeqn
and
\eqn
        &&\int_{(\Tor^3)^{\bar n-q}}\Big[\prod_{r\in T^c}
        \;dp_r\; \frac{1  }
        {|\en(p_r)-\alpha_r-i\sigma_r\e|^{\mu_j(r)}}\Big]
	\nonumber\\
	&&\hspace{3cm}\times
        \prod_{s\in T}
        \frac{1}{|\en(w_s)-\alpha_s-i\sigma_s\e|^{\mu_j(s)}}
        \nonumber\\
        &\leq& (C|\log\e|)^{\bar n-q}
        \e^{-\sum_{r\in T^c}\delta_j,r-\sum_{s\in T}\mu_j(s)} \;,
\eeqn
we find
\eqn
        |\amp|&\leq&\lambda^{2\bar n}(C  |\log \e|)^{\bar n-q+2}
        \e^{1-\mu_j(n)-\sum_{r\in T^c} \delta_{j,r}
        -\sum_{s\in T} \mu_j(s)} (C \e^{-1})^q \e^{3/2}
        \nonumber\\
        &\leq&(C\lambda^2\e^{-1}|\log\e|)^{\bar n}\e^{1/2} \; ,
\eeqn
due to
$$
        \mu_j(n)+\sum_{r\in T^c} \delta_{j,r}+\sum_{s\in T}
        \mu_j(s) =\bar n-q+2 \;,
$$
where $q\geq2$. This proves the lemma.
\endprf

\section{Type III contractions}

We recall that the number of type III contractions $\pi\in\Pi_{n,n'}$
is superfactorially large, bounded by $\bar n^{2\bar n}$.
On the other hand, if $\pi$ is of type III, Lemma {~\ref{highercorrlm}} below
shows that $|\amp|$ is by some positive powers of $\e$ smaller than the
bounds on crossing or nesting pairing graphs.
This will suffice to balance the extremely large combinatorial factors
against the size of $|\amp|$.

\begin{lemma}\label{highercorrlm}
Assume that $1\leq m< \bar n=\frac{n+n'}{2}\in\N$,
and $\pi=\{ S_j\}_{j=1}^m\in\Pi_{n,n'}$ of type
III. Then, for $m=\bar n-1$,
\eqn
        |\amp| \leq     \e|\log \e|^3
                 (C\e^{-1}\lambda^2|\log \e|)^{\bar n}\;,
\eeqn
while for all $1\leq m\leq \bar n -2$,
\eqn
                 |\amp| \leq (c\bar n)^{3\bar n-1} \lambda^{2\bar n} \e^{-m}
                 (C|\log \e|)^{2\bar n} \;.
\eeqn
\end{lemma}

\prf
By assumption, the total number of blocks contained
in the contraction $\pi$ is $m$, and we recall that the
case $m=\bar n$ excluded here would correspond to a pairing graph.
After integrating out $\delta(p_n-p_{n+1})$,
\eqn
        |\amp|&\leq&\frac{\lambda^{2\bar n} e^{2\e t}}{(2\pi)^2}
        \int_{I\times \bar I} |d\alpha| \,|d\beta| \int_{\Tor^3} dp_n\;
        \frac{1}{|\en(\vp_n)-\alpha-i\e|}
        \,\frac{1}{|\en(\vp_n)-\beta-i\e|}
        \nonumber\\
        &\times&
        \int_{(\Tor^3)^{2n}}dp_0\cdots dp_{n-1}dp_{n+2}\dots dp_{2\bar n+1}
        \nonumber\\
        &\times&
        \Big[\prod_{j=0}^{n-1}\frac{1}{|\en(p_j)-\alpha-i\e|}\Big]
        \prod_{\ell=n+2}^{2n+1}\frac{1}{|\en(p_\ell)-\beta+i\e|}
        \nonumber\\
        &\times&\prod_{ j=1 }^m
        \big|c_{|S_j|}\big|\delta_{S_j}(\up)
          \;. \; \;
\eeqn
Let $J:=\sharp\{j|\,|S_j|>2\}$ denote the number of type III blocks
in $\pi$, hence the number of pairings is $m-J$.

We consider the graph $G_\pi$ associated to the contraction $\pi$.
$G_\pi$ contains one vertex from $\delta(p_n-p_{n+1})$, $2\bar n$
vertices corresponding to $V_\omega$, $J$ vertices in $V_{hc}(G_\pi)$
(cf. the definition in the second paragraph of section
{~\ref{graphsection}}), and we add
two artificial vertices at the free ends of the particle lines
corresponding to the initial conditions
(labelled by the momenta $p_0$ and $p_{2\bar n+1}$).

Every type III block $S_j$ accounts for $|S_j|$ contraction lines,
while for a pairing block, there is only $\frac{|S_j|}{2}=1$
contraction line.
The total number of contraction lines in $G_\pi$ is thus
\eqn
        \sum_{j\in J}|S_j|+(m-J)=2\bar n-(m-J) \;,
\eeqn
(since $\sum_{j\in J}|S_j|+2(m-J)=2\bar n$ is the total number of
$V_\omega$-vertices).

Let $T$ denote a spanning tree of $G_\pi$ which contains all
contraction lines,
the two particle lines belonging to the momenta $p_n$ and $p_{2\bar n+1}$,
but not the particle line that used to belong to $p_{n+1}$. Clearly,
$T$ has $2\bar n+2+J$ edges, from which $(2\bar n+2+J)-2-(2\bar n-(m-J))=m$ belong to
particle lines different from those labelled by $p_n$ and $p_{2\bar n+1}$.
All particle momenta associated to those particular edges of $T$ can
be expressed
as linear combinations of momenta not on $T$ (they are used to
integrate
out all delta distributions).
Accordingly, we estimate all propagators on $T$ except for those
labelled by $p_n$ and $p_{2\bar n+1}$ by their $L^\infty$-norms.
This yields a factor $\e^{-m}$.

We integrate the propagators labelled by $p_n$ and $p_{2\bar n+1}$ against
$\alpha$ and $\beta$, respectively, which yields a factor
$(c\log\frac1\e)^2$.
Furthermore, we bound all $2\bar n-m$ remaining propagators that belong
to edges in the complement of $T$ by their $L^1$-norms.
This produces a factor $(c\log\frac1\e)^{2\bar n-m}$.

Finally, we derive from  (~\ref{renormc2kbound}) that
\eqn
        \prod_{ j=1 }^m
        \big|c_{|S_j|}\big|&\leq&
        \prod_{j=1}^m
        (|S_j|\,e^{\frac12 e^{\com}})^{|S_j|+1}
        \nonumber\\
        &\leq&
        (2\bar n \, e^{\frac12 e^{\com}} )^{3\bar n-1} \;
\eeqn
for $1\leq m\leq \bar n-2$, so that in this case,
$$
        |\amp| \leq (C\bar n)^{3\bar n-1} (C\lambda^2)^{\bar n}  \e^{-m}
        |\log \e|^{2\bar  n-m+2} \;.
$$
On the other hand, since $c_2=1$ by normalization,
\eqn
        \prod_{ j=1 }^{\bar n-1}
        \big|c_{|S_j|}\big|=| c_4 | \;
\eeqn
if $m=\bar n-1$, so that
$$
        |\amp| \leq | c_4 | (C\lambda^2)^{\bar n}  \e^{-\bar n+1}
        |\log \e|^{\bar n+3} \;.
$$
This proves the lemma.
\endprf

\begin{proposition}
For fixed $n,n'$, the sum of all contributions to the expectation
(~\ref{exppotgenexpr}) that comprise type III contractions is bounded
by
$$
        \sum_{\pi\in\Pi_{n,n'}\; {\rm type \;III}}|\amp|
        \leq    (C\lambda^2 \e^{-1}|\log\e|^2 )^{\bar n} \Big(
        \bar n^4(\bar n!)\e  +  \bar n^{5\bar n} \e^2 \Big) \;.
$$
\end{proposition}

\prf
We note that the total number of graphs for $1\leq m\leq \bar n-2$ is
bounded by
$$
        \sum_{m=1}^{\bar n-2}B_{\bar n}(m)
        <   \bar n^{2\bar n+1} \;,
$$
cf. the discussion of (~\ref{Bmdef}). In the case $m=\bar n-1$, we find
$$
        B_{\bar n}(\bar n-1) < 2^{\bar n} (\bar n!) \bar n^4 \;,
$$
since we have $\bar n-1$ pair
correlations, and one correlation of order 4.
Application of Lemma {~\ref{highercorrlm}} implies the claim.
\endprf

\section{Estimates on the remainder term}

In this section, we bound the expectation of the $L^2$-norm
of the remainder term $\Rem_{N,t}$ in the Duhamel series (~\ref{Duh-def-1}).
We shall use the partial time integration method introduced in \cite{erdyau}.

The remainder term is defined by
\eqn
        \Rem_{N,t} =-i\lambda\int_0^t ds \, e^{-i(t-s)H_\omega}
        V_\omega\phi_{N,s} \;.
\eeqn
Let $\kappa\in\N$, $1\ll\kappa\ll N$, be a large integer
to be chosen later. We subdivide $[0,t]$ into $\kappa$ subintervals
with equidistant boundary points $\{\tt_0,\dots,\tt_{\kappa}\}$ where
$t_0=0$, $\tt_{\kappa}=t$, such that
\eqn
        \Rem_{N,t} =-i\lambda\sum_{j=0}^{\kappa-1}
        e^{-i(t-\tt_{j+1})H_\omega}
        \int_{\tt_j}^{\tt_{j+1}} ds \,
        e^{-i(\tt_{j+1}-s)H_\omega} V_\omega\phi_{N,s}
        \; .
        \label{remainderterm}
\eeqn
Furthermore, we define
$$
        \hat\phi_{m,n,\tt}(s) := \int_\tt^{s}d s'
        D_{s-s'}^{(m-n)}\hat\phi_{m,s'} \;,
$$
where
\eqn
        &&(D_{t}^{(m)}\hat\phi)(\vp_0) := (-i\lambda)^m
        \int_{\R^{n+1}}\Big[\prod_{j=0}^m ds_j\Big]_t
        e^{-i s_0 \en(\vp_0)}
        \\
        &&\hspace{1cm} \times\,
        \int_{(\Tor^3)^m} d\vp_1\cdots d\vp_m  \Big[ \prod_{j=1}^m
        e^{-i s_j \en(\vp_j)} \hat V_\omega
        (\vp_j-\vp_{j-1})\Big]\hat\phi_{s_m}(\vp_m)\; .
        \nonumber
\eeqn
$\hat\phi_{m,n,\tt}(s)$ is the $m$-th Duhamel term, comprising
$m$ collisions in total
with $V_\omega$, but conditioned on the requirement that
precisely $n$ collisions occur before time
$\theta$.

We then split the remainder term into
\eqn
        \Rem_{N,t}&=&\Rem_1(t)+\Rem_2(t)
\eeqn
with
\eqn
            \Rem_1(t)&:=&-i\lambda\sum_{N\leq n<4N}\sum_{j=0}^{\kappa-1}
            e^{-i(t-\tt_{j+1})H_\omega} V_\omega
            \phi_{n,N,\tt_j}(\tt_{j+1}) \;,
            \\
            \Rem_2(t)&:=&-i\lambda \sum_{j=0}^{\kappa-1}
            e^{-i(t-\tt_{j+1})H_\omega} \int_{\tt_j}^{\tt_{j+1}}ds \,
            e^{-i(\tt_{j+1}-s)H_\omega } V_\omega\phi_{4N,N,\tt_j}(s) \; .
            \nonumber
\eeqn
$R_1(t)$ is obtained from further expanding the operators
$e^{-i(\theta_{j+1}-s)H_\omega}$ in (~\ref{remainderterm}) up to $3N-1$
times.
Consequently, $R_1(t)$ comprises Duhamel terms for which up to $3N-1$
collisions occur
in a time interval of length $\frac{t}{\kappa}$. $R_2(t)$ is the
corresponding
error term, characterized by the fact that precisely $3N$ collisions
occur in a time interval of that length.

Our aim is to establish that $\Exp[\|\Rem_{1,2}(t)\|_2^2]=O(\e^\delta)$
for
some $\delta>0$. The estimates used to control
$\Exp[\|\Rem_1(t)\|_2^2]$ are essentially equal to those employed for
$n\leq N$.
To bound $\Exp[\|\Rem_2(t)\|_2^2]$, we exploit the rarity of events
comprising large collision numbers (of order $O(N)$) in the time
intervals $[\tt_{j},\tt_{j-1})$ that are much shorter
than $[0,t]$.

\begin{lemma}
There are finite constants $C$, uniform in $\e=t^{-1}$ and $N$, such that
\eqn
        \Exp\Big[\| \Rem_1(t)\|_2^2\Big]
        &\leq& \frac{N^2\kappa^2 (C\lambda^2 \e^{-1} )^{4N }}{(N!)^{1/2}}
        \label{ExpRem1est}\\
        &+& N^2 \kappa^2 (C\lambda^2 \e^{-1}|\log\e|)^{4N} |\log \e|^{3}
        \Big( \e^{\gex }(4N)! + \e^2 (4N)^{20N}\Big)
        \nonumber\\
        &&\nonumber\\
        \Exp\Big[\| \Rem_2(t) \|_2^2\Big]
        &\leq& \e^{-2}  (C\lambda^2 \e^{-1}|\log\e|)^{4N} |\log\e|^{3}
        \label{ExpRem2est}\\
        &\times&
        \Big(\kappa^{-N} (4N)!+\kappa^{-N+5}\e (4N)! (4N)^4
        \nonumber\\
        &&\;\;\;\;\;\;
        +\kappa^{-N+9}\e^2(4N)! (4N)^8
        +\e^3(4N)^{20N} \Big) \; .
                \nonumber
\eeqn
\end{lemma}

\prf
The Schwarz inequality and unitarity of $e^{-it H_\omega}$ imply
\eqn
        \Exp\Big[\|\Rem_1(t)  \|_2^2\Big] &\leq&
        (3N)^2\kappa^2 \sup_{N<n\leq4N} \sup_{0\leq j<\kappa}
        \Exp\Big[\| \phi_{n,N,\tt_j}(\tt_{j+1}) \|_2^2  \Big]
        \nonumber\\
        \Exp\Big[\|\Rem_2(t) \|_2^2\Big] &\leq& \e^{-2}
        \sup_{\stackrel{0\leq j<\kappa}{s\in[\tt_j,\tt_{j+1}]}}
        \Exp\Big[\| \phi_{4N,N,\tt_j}(\tt_{j+1}) \|_2^2 \Big] \;.
        \label{R2est}
\eeqn
Let us first address the estimates on $\|\Rem_2(t)\|_2^2$.

Each pairing contraction occurring in
$\Exp[\| \phi_{4N,N,\tt_j}(\tt_{j+1}) \|_2^2]$
can be bounded by
\eqn
        \kappa^{-N} |\log\e|^3 (C \lambda^2 \e^{-1}|\log\e|)^{4N}  \;.
\eeqn
The factor $\kappa^{-N}$ appears for the following reason.
We recall that $\e^{-1}=t$ is the length
of the time integration interval $[0,t]$,
and that previously, $i\e$ has appeared as the imaginary part of the
denominators of the
free resolvents in the momentum space Feynman integrals.
Due to the condition in $R_2(t)$ that
all of the last
$3N$ collisions occur in a time interval of length
$\frac t\kappa\ll t$,
there are $6N$ out of $2(4N+1)$ free resolvents,
for which the imaginary part of the denominator is $i\kappa\e$
instead of $i\e$.
$i\e$ appears only in $2N+2$ of the
free resolvents, corresponding to the first $N$ collisions.

For type III contractions, we argue as in the proof of
Lemma {~\ref{highercorrlm}}. We observe that if there is a
single block of size 4 (that is, one delta contracting
4 random potentials), we gain a factor $\e$, and there
are $2(4N+1)-4$ free resolvents which are part of pairing contractions.
The above considerations apply to the latter, and there
is a gain of a factor of at least $\kappa^{-N+5}$. The number
of type III contractions with only one block of size 4
is bounded by $(4N)^4(4N)!$.

For a type III contraction which contains two size 4 blocks or one
size 6 block, we gain a factor $\e^2$, and there
are at least $2(4N+1)-8$ free resolvents which are part of pairing contractions.
By the above, we gain a factor of at least $\kappa^{-N+9}$. The number
of type III contractions with two blocks of size 4 or one block
of size 6 is bounded by $(4N)^8(4N)!$.

Any type III contraction with larger or more non-pairing
blocks provides a gain of a factor $\e^3$, and we shall
then not need inverse powers of $\kappa$. The number of
such contractions, multiplied with the estimate derived in the
proof of Lemma {~\ref{highercorrlm}}
on the renormalized moments, is bounded by $c^N(4N)^{20N}$.

Hence, we conclude that
\eqnn
            \Exp\Big[\| \phi_{4N,N,\tt_j}(\tt_{j+1}) \|_2^2 \Big]
            &\leq&|\log\e|^3 (C\lambda^2 \e^{-1}|\log\e|)^{4N}
            \\
            &\times&
            \Big(\kappa^{-N}(4N)!  \kappa^{-N+5}\e (4N)! (4N)^4
            \nonumber\\
            &&\;\;\;\;\;\;
            +\kappa^{-N+9}\e^2 (4N)! (4N)^8
            +\e^3(4N)^{20N} \Big) \;,
\eeqnn
where the first term on the right hand side of the inequality sign
stems from the sum over all pairing contractions,
while the second term accounts for all type III contractions.
This proves the asserted estimate on $\Exp[\|\Rem_2(t)\|^2]$.
A more detailed exposition is given in \cite{erdyau}.

The bound on $\Exp[\|\Rem_1(t)\|_2^2]$ follows from Lemmata
{~\ref{partintsimppairlm}}, {~\ref{crossgeqN}}, {~\ref{nestgeqN}},
{~\ref{partinthighcorrlm}} below.
\endprf

\subsection{Pairing contractions}

Let us first estimate the contributions to $\Exp[\|R_1(t)\|^2]$
stemming from pairing contractions.
For simple and crossing pairings, the
necessary bounds on terms corresponding to $n$ with
$N<n\leq 4N$ are precisely the same as for
$n\leq N$. The discussion of nested pairing
contractions is slightly more involved, due to the fact that particle
propagators with different imaginary parts $\pm i\e$ and $\pm
i\kappa\e$
can appear in the same simple nest.

\begin{lemma}\label{partintsimppairlm}
Let $N<n\leq 4N$, and $\lambda^2\e^{-1}<1$.
The contribution to (~\ref{ExpRem1est})
of the sum of all simple pairings is bounded by
\eqn
            \Big|\sum_{\pi\in\Pi_{n,n}\;{\rm simple}}\amp\Big|\leq
            \frac{(C_0\lambda^2 \e^{-1})^{ n}}{(n!)^{1/2}} +
            n \e^{\gex } |\log\e|^3
            (C \e^{-1} \lambda^2|\log\e|)^{n} \;,
\eeqn
where $C_0$ is defined in (~\ref{c0def}).
\end{lemma}

\prf  The proof is derived from the same arguments as in the proof of
Lemma {~\ref{simplepairlemmleqN}}. Here,
\eqn
            \frac{1}{|\en(p_j)-\alpha_j-i\sigma_j\kappa\e|}\leq
            \frac{1}{|\en(p_j)-\alpha_j-i\sigma_j\e|}
            \label{resreskappaest}
\eeqn
isw used for all $j$. \endprf

The remark after the proof of
Lemma {~\ref{simplepairlemmleqN}} concerning globality in
$T=\lambda^2t>0$
also applies to the present situation.

\begin{lemma}\label{crossgeqN}
Let $N<n<4N$, and let $\pi\in\Pi_{n,n}$ correspond to a pairing
contraction that
contains at least one crossing. Then,
$$
            |\amp|\leq \e^{\gex }|\log\e|^3
            (C\lambda^2\e^{-1}|\log\e|)^n \;.
$$
\end{lemma}

\prf
The proof is analogous to that of Lemma
{~\ref{crosslessN}},
and uses (~\ref{resreskappaest}).
\endprf

\begin{lemma}\label{nestgeqN}
Let $N<n<4N$, and let $\pi\in\Pi_{n,n'}$ represent a non-crossing
pairing contraction that contains at least one nested subgraph. Then,
$$
            |\amp|\leq  \e^{\frac12}|\log\e|^3(C\lambda^2\e^{-1}|\log\e|)^n   \;.
$$
\end{lemma}

\prf
In the case $N<n<4N$, particle resolvents with imaginary parts
$i\e$ and $i\kappa\e$ in the denominator can appear simultaneously in
the same nested pairing subgraph.
If so, $\amp$ contains a subintegral corresponding to a
nest of the form
\eqn
        N_{q_1,q_2 }(\alpha,\e,\kappa)\delta(p_{i+2q-1}-p_{i-1}) 
        &:=&\lambda^{2q}\int_{(\Tor^3)^{2q}}   dp_{i} \cdots dp_{i+2q-2}
        \delta(p_i-p_{i-1}+p_{i+2q-1}-p_{i+2q-2})
        \nonumber\\
        &\times&
        \prod_{j=1}^{q-1} \delta(p_{i+2j+1}-p_{i+2j-1})
        \nonumber\\
        &\times&
        \Big(\prod_{l=i}^{N}\frac{1}{\en(p_l)-\alpha-i\e}\Big)
        \prod_{k=N+1}^{i+2q-2}\frac{1}{\en(p_k)-\alpha-i\kappa\e}
        \;,\label{Nq1q2kappadef}
\eeqn
where for $q_1+q_2=q-1$, and $q_1,q_2\geq1$,
\eqn\label{Nq1q2alphedef}\\
        \nonumber
        N_{q_1,q_2  }(\alpha,\e,\kappa) =\int_{\Tor^3} dp_i
        \frac{\, \lambda^2 \, (\lambda^2\Th(\alpha,\e))^{q_1}}
        {(\en(p_i)-\alpha-i\e)^{q_1+\th}}
        \frac{(\lambda^2\Th(\alpha,\kappa\e))^{q_2} }
        {(\en(p_i)-\alpha-i\kappa\e )^{q_2+1-\th}} \;,
\eeqn
with $\th =0$ or $1$. Let us assume that $\th=0$, the case $\th=1$ is
completely
analogous. Then,
\eqn
    (~\ref{Nq1q2alphedef})&=&\lambda^2(\lambda^2\Th(\alpha,\e))^{q_1}
        (\lambda^2\Th(\alpha,\kappa\e))^{q_2}
        \nonumber\\
        &\times&
        \int_{\Tor^3}dp_i\Big( \frac{1}{(q_1-1)  !}
        \partial_{\alpha}^{q_1-1  }
        \int_{\R_+ } ds_1
        e^{-is_1(\en(p_i)-\alpha-i\e)}
        \nonumber\\
        &\times&
        \frac{1}{q_2 !}\partial_{\alpha}^{q_2  }
        \int_{\R_+} ds_2
        e^{-is_2(\en(p_i)-\alpha-i\kappa\e)} \Big)
        \nonumber\\
        &=& \lambda^2\frac{1}{(q_1-1)  !}
        \frac{1}{q_2 !}(\lambda^2\Th(\alpha,\e))^{q_1}
        (\lambda^2\Th(\alpha,\kappa\e))^{q_2}
        \\
        &\times&
        \int_{\R_+^2 } ds_1 ds_2 (is_1)^{q_1-1  }  (is_2)^{q_2 }
        \nonumber\\
        &\times&
        \int_{\Tor^3}dp_i
        e^{-i(s_1+s_2)(\en(p_i)-\alpha)}e^{-\e s_1-\kappa\e s_2}
        \;.\nonumber
\eeqn
Using (~\ref{deralmThinf}), the integral on the last line is bounded by
\eqn
        &&\int_{\R_+^2 } ds_1 ds_2 s_1^{q_1-1 }  s_2^{q_2  }
        \frac{C^{q_1+q_2}}{(1+s_1+s_2)^{\frac32}}
        \;e^{-\e s_1-\kappa\e s_2}
        \nonumber\\
        &\leq& \frac{\e^{-(q_1+q_2-1/2)}}{\kappa^{q_2  }}
        \int_{\R_+^2 } ds_1 ds_2
        s_1^{q_1 -1 }  s_2^{q_2  }
        \frac{C^{q_1+q_2}}{(1+s_1)^{\frac32}}\;e^{- s_1- s_2}
        \nonumber\\
        &<& \frac{\e^{-(q-3/2)}}{\kappa^{q_2  }}\;C^{q_1+q_2}\;
        ((q_1 -1)!)(q_2!) \;.
\eeqn
Therefore,
\eqn
        |N_{q_1,q_2  }(\alpha,\e,\kappa)|
        <
        \frac{\e^{3/2} (C \lambda^2\e^{-1})^{q}}{\kappa^{q_2  }} \;,
\eeqn
where (~\ref{deralmThinf}) has been used.
We note that in the special case $q_1=0$, $q_2=q-1$, this
is replaced by
\eqn
        |N_q(\alpha,\kappa\e)| \leq
        \frac{(C\e^{-1}\lambda^2)^q\e^{-3/2} }{\kappa^{q-3/2}} \;,
\eeqn
cf. (~\ref{Nqalphaeest}). For the assertion of this lemma, it
is, however, not necessary to
take advantage of the small inverse powers in $\kappa$.

For the contractions outside of the nest, we proceed as in the
proof of Lemma {~\ref{nestlessN}} (where in (~\ref{Nq1q2kappadef}),
$i:=j+1$),
and find
\eqn
        |\amp|\leq\e^{\frac12}
        |\log\e|^4(C \lambda^2\e^{-1}|\log\e|)^n\;.
\eeqn
This proves the lemma.
\endprf

\subsection{Type III contractions}

The estimates on type III contractions necessary for $n>N$ are the same
as for $n\leq N$.

\begin{lemma}\label{partinthighcorrlm}
Let $N<n\leq 4N$, and let $\pi\in\Pi_{n,n}$ correspond to a type III
contraction. Then,
$$
        \sum_{\pi\in\Pi_{n,n}\;{\rm type \; III}}|\amp|\leq
        (C\lambda^2 \e^{-1} |\log\e| )^{n} \Big( (n!)\e  +
        n^{5n} \e^2 \Big) \;,
$$
where the constant $C$ is uniform in $\e$, $\lambda$, and $n$.
\end{lemma}

\prf
This is proved in the exact same way as Lemma {~\ref{highercorrlm}}.
We remark that subfactorial factors
$n^4$, $4^n$, etc. have here been absorbed into the multiplicative
constant.
\endprf

\section{Completion of the proof}
\label{mainlemmaprf}

Collecting the above, we are in the position now to prove the key
estimate (~\ref{fundest10}), which concludes the proof of
Lemma {~\ref{fundestlemma}}.
Combining Lemmata {~\ref{simplepairlemmleqN}}, {~\ref{crosslessN}},
{~\ref{nestlessN}}, we find
\eqn
                | l.h.s. \;of\; (~\ref{fundest10})|
                &\leq&
                C_1  \lambda^2 \e^{-1}
                \nonumber\\
                &+&
                (4N\kappa)^2  |\log\e|^4
                (C\lambda^2\e^{-1}|\log \e|)^{4N}
            	\Big[\e^{\gex }(4N)! +  \e^2(4N)^{20N}\Big]
                \nonumber\\
                &+&\e^{-2}
                |\log\e|^3  (C\lambda^2 \e^{-1}|\log\e|^2 )^{4N}
            	\nonumber\\
            	&&\hspace{2cm}\times\;
                \Big[\kappa^{-N} (4N)!+\kappa^{-N+5}\e (4N)! (4N)^4
            	\nonumber\\
            	&&\hspace{5cm}
            	+\kappa^{-N+9}\e^2(4N)! (4N)^8
            	+\e^3(4N)^{20N} \Big]\;,
\eeqn
where some subexponential factors, such as $N^4$, etc., have been
absorbed
into the multiplicative constants $C$ in $C^{4N}$,
and where the constant $C_1$ is defined in (~\ref{c0def}).

According to the assumptions of Lemma {~\ref{fundestlemma}}, we have
$$
        \e^{-1}=t=\delta^{\frac67}\lambda^{-2}\;,
$$
where $0<\delta<1$ is given and fixed.

Furthermore, we choose
\eqn
        N(\e)&=& \lfloor\;\frac{|\log\e|}{40\log|\log\e|}
        \;\rfloor  \; ,
        \nonumber\\
        \kappa(\e)&=&\lceil\;|\log\e|^{120} \; \rceil \;.
\eeqn
One then easily verifies that
\eqn
        (4N(\e))^{20N(\e)}&<&\e^{-\frac{1}{2}}
        \nonumber\\
        (4N(\e))!&<&\e^{-\frac{1}{10}}
        \nonumber\\
        \kappa(\e)^{N(\e)}&\sim&\e^{-3} \;,
\eeqn
such that for instance,
\eqn
          \e^{-2}\kappa(\e)^{-N(\e)} ((4N(\e))!)<\e^{1/2} \;.
\eeqn
It can then be straightforwardly verified that for $\e$ sufficiently small,
\eqn
        | l.h.s. \; of  \; (~\ref{fundest10})|&\leq&
        C_1 \delta^{\frac67}+\e^{\frac{1}{7}} \;.
\eeqn
This completes the proof of Lemma {~\ref{fundestlemma}}.

\section{Linear Boltzmann equations}

We shall in this section study the Schr\"odinger
dynamics of the random lattice model analyzed above, and demonstrate
that its macroscopic, weak coupling limit is governed by
the linear Boltzmann equations.  Owing to the similarity of many
of the arguments used here to those presented in \cite{erdyau} for
the continuum case, the exposition will be very condensed.

Let $\phi_t\in \ell^2(\Z^3)$ denote the solution of the random
Schr\"odinger equation
\eqn\label{RSE}
        i\partial_t\phi_t=H_\omega \phi_t \;,
\eeqn
with initial condition $\phi_0\in \ell^2(\Z^3)$, and for a fixed
realization of the
random potential.
We define its Wigner transform
$W_{\phi_t}:(\Z/2)^3\times\Tor^3\rightarrow\R$ by
\eqn
        W_{\phi_t}(x,v)=8\sum_{y,z\in\Z^3\atop y+z=2x}\overline{\phi_t(y)}\phi_t(z)
        e^{2\pi iv(y-z)} \;,
\eeqn
where we note that $x\in(\Z/2)^3$.
Fourier transformation with respect to $x$ yields
\eqn
        \hat W_{\phi_t}(\xi,v)=
        \overline{\hat\phi_t(v-\xif)}\hat\phi_t(v+\xif)\;,
\eeqn
where $v\in\Tor^3$ and $\xi\in(2\Tor)^3$.

Let $J$ denote a Schwartz class function on $\R^3\times \Tor^3$.
We introduce macroscopic time, space, and velocity
variables $(T,X,V):=(\eta t,\eta x,v)$ for $\eta\ll1$,
and the rescaled, macroscopic Wigner transform of $\phi_t$
\eqn
        W^{(\eta)}_{T}(X,V):=\eta^{-3}W_{\phi_{T/\eta}}(X/\eta,V) \;,
\eeqn
with $X\in(\eta\Z/2)^3$, $V\in\Tor^3$.
Then, let
\eqn
        \langle J,W^{(\eta)}_{T}\rangle &=&
        \sum_{X\in(\eta\Z/2)^3}\int_{\Tor^3}dV
        \overline{J(X,V)}W^{(\eta)}_{\phi_{T/\eta}}(X,V)\; ,
        \label{eqn3-1-0}
\eeqn
while for the Fourier transform with respect to the first argument,
\eqn
        \langle J,W^{(\eta)}_{T}\rangle= \langle \hat J,
        \hat W^{(\eta)}_{\phi_t}\rangle =
        \int_{(2\Tor)^3\times\Tor^3 }d\xi dv
        \overline{\hat J_\eta(\xi,v)}
        \hat W_{\phi_t}(\xi,v)  \; ,
        \label{eqn3-1-1}
\eeqn
where $\hat J_\eta(\xi,v):=\eta^{-3}\hat J(\xi/\eta,v)$.

We shall write
$\sin 2\pi w\in[-1,1]^3$ for the 3-vector with components
$ \sin 2\pi w_j$, $j=1,2,3$, where $w\in\Tor^3$.

\begin{theorem}\label{Boltzlimthm}
Let the scaling factor be fixed by 
\eqn
        \eta=\lambda^2 \;,
        \label{eta-lamb-cond-1}
\eeqn
where $\lambda$ is the disorder strength.
Let $\phi_t^{(\eta)}=e^{-itH_\omega}\phi_0^{(\eta)}$
denote the solution of the random Schr\"odinger equation
(~\ref{RSE}) with initial condition
\eqn
        \phi_0^{(\eta)}(x)=\eta^{3/2} h(\eta x)
        e^{i s(\eta x)/\eta} \;,
\eeqn
where $h,s$ are Schwartz class functions on $\R^3$.

Let $W_T^{(\eta)}$ denote the rescaled, macroscopic
Wigner transform of $\phi_t^{(\eta)}$.
Then, for any $T>0$, it has the weak limit
\eqn
        w-\lim_{\eta\rightarrow0}\Exp\big[ W^{(\eta)}_{T}(X,V)\big]= F_T(X,V)\;,
\eeqn 
where $F_T(X,V)$ solves the linear Boltzmann equation
\eqn
        &&\partial_T F_T(X,V) + \sin2\pi V \cdot
        \nabla_X F_T(X,V)
        \nonumber\\
        &&\hspace{2cm}=  \int_{\Tor^3} dU   \sigma(U,V)
        \lb F_T(X,U) - F_T(X,V)\rb \;,
        \label{linB}
\eeqn
with collision kernel
\eqn
        \sigma(U,V)=2\pi\delta(\en(U)-\en(V)) \;,
\eeqn
and initial condition
\eqn\label{initcondweaklim}
        F_0(X,V)
        =w-\lim_{\eta\rightarrow0}W_{0}^{(\eta)}
        =|h(X)|^2\delta(V-\nabla s(X)) \;.
\eeqn
\end{theorem}

\prf
For (~\ref{initcondweaklim}), we refer to \cite{erdyau}.

Let
$$
        \phi^{main}_{t}:=\sum_{n=0}^N \phi_{n,t}\;,
$$
and
\eqn
        \Exp \lb \int_{\Tor^3\times\Tor^3}d\xi dv
        \overline{\hat J_\eta(\xi ,v)}
        \hat W_{\phi_t^{main}}(\xi,v)\rb
        &=&\sum_{n,n'=0}^N U^{\hat J_\eta}_{n,n'}\nonumber\\
        &=&\sum_{n,n'=0}^N \sum_{\pi\in \Pi_{n,n'}}
        \ampl_{\hat J_\eta}[\pi] \;,
        \label{WigweakDuhamel}
\eeqn
where
$$
        U_{n,n'}^{\hat J_\eta}=\Exp \lb \int_{\Tor^3\times\Tor^3}d\xi dv
        \overline{\hat J_\eta(\xi ,v)}\;\overline{\hat \phi_{n,t}(v-\xif)}
        \hat\phi_{n',t}(v+\xif) \rb\;.
$$
$\ampl_{\hat J_\eta}[\pi]$ denotes the value of the
integral corresponding to the contraction $\pi\in\Pi_{n,n'}$, .

\begin{lemma}
Let $\pi\in\Pi_{n,n'}$, and $\bar n:=\frac{n+n'}{2}\in\N$. Then,
\eqn
        U_{n,n'}^{\hat J_\eta} &=&
        \sum_{\pi \in \Pi_{n,n'}\;{\rm simple}}
        \ampl_{\hat J_\eta}[\pi]
        \nonumber\\
        &+& O\Big( (C\lambda^2 t\log t)^{\bar n}  (\log t)^{3}
        \big(t^{-\gex } \bar n ! +
        t^{-2}\bar n^{5\bar n}\big)\Big) \;,
        \label{WigweakDuhest}
\eeqn
and for any simple pairing $\pi$,
\eqn
        |\ampl_{\hat J_\eta}[\pi]| \leq (C\lambda^2 t)^{\bar n} \;.
\eeqn
\end{lemma}

\prf
The only difference here in comparison to the
$L^2$-bounds previously considered is
the presence of $J_\eta$.
We note that for the choice $J_\eta=\delta(\xi)$, (~\ref{WigweakDuhamel})
reduces to
$\Exp[\|\phi_t^{main}\|_{\ell^2(\Z^3)}^2]$ as treated earlier.
The necessary modifications are straightforward, and the same
estimates on  Feynman amplitudes enter as before,
and as expressed in (~\ref{WigweakDuhest}).
For a detailed account on these matters,  we refer to \cite{erdyau}.
\endprf

Let $\e=\frac1t$, as before.
Similarly as in the proof of Lemma {~\ref{simplepairlemmleqN}},
we decompose $\ampl_{\hat J_\eta}[\pi]$, for $\pi$ simple, into a
main part $\ampl_{\hat J_\e,main}[\pi]$, and an error part, where
$\ampl_{\hat J_\eta,main}[\pi]$ is obtained by replacing the recollision terms
$\Th (\alpha,\e)$ and
$\Th (\beta,-\e)$ in $\ampl_{\hat J_\eta}[\pi]$ by $\Th (\en(v_0),\e)$ and
$\Th (\en(v_0),-\e)$.
We assume for $\pi$ that $\ampl_{\hat J_\eta}[\pi]$ contains $m$ type
II contractions,
where we index the immediate recollisions by
$(q_0,\dots,q_m)$ and $(\tilde q_0,\dots,\tilde q_m)$, respectively,
as in (~\ref{simplepairfixedn}). Then, we have
\eqn
        \ampl_{\hat J_\eta,main}[\pi]
        &=& \frac{ \lambda^{2m}e^{2\e t} }{(2\pi)^2}
        \int_{I\times \bar I} d\alpha d\beta
        e^{-it(\alpha-\beta)}
        \int_{\Tor^3\times\Tor^3} d\xi dv_0
        \overline{\hat J_\eta( \xi,v_0)}
        \nonumber\\
        &\times&
        \int_{(\Tor^3)^m} dv_1\cdots dv_m
        \overline{\phi_0^{(\eta)}(v_n-\xif)}
        \phi_0^{(\eta)}(v_n+\xif)
        \\
        &\times&\prod_{i=0}^m
        \frac{(\lambda^2\Th(\en(v_0),\e))^{q_i} }
        {(\en(v_i+\xif)-\alpha-i\e)^{q_i+1}} \,
        \frac{(\lambda^2\Th(\en(v_0),-\e))^{\tilde q_i}}
        {(\en(v_i-\xif)-\beta+i\e)^{\tilde q_i+1}}
        \;.\nonumber
\eeqn
The error term is controlled by the following lemma.

\begin{lemma}
Let $\pi\in\Pi_{n,n'}$ be a simple pairing. Then,
\eqn
        \ampl_{\hat J_\eta}[\pi]&=&\ampl_{\hat J_\eta,main}[\pi]
        +O((C\lambda^2 t)^{\bar n} t^{-\gex })
        \nonumber\\
        |\ampl_{\hat J_\eta,main}[\pi]|&\leq&
        \frac{(C\lambda^2 t)^{\bar n}}{(\bar n !)^{1/2}} \;.
\eeqn
\end{lemma}

\prf
The proof is analogous to the one of Lemma {~\ref{simplepairlemmleqN}},
with straightforward modifications to accommodate for $\hat J_\eta$.
This is treated in detail for the continuum model in \cite{erdyau},
and we shall not reiterate it here.
\endprf

We perform the contour integral with respect to the variables $\alpha$
and $\beta$, and evaluate the sum over $n,n'\in\{0,\dots,N\}$ by first
summing over all $q_i,\tilde q_i$, where $i=1,\dots,m$,
for fixed $m$, and subsequently summing over the indices $m$.
We then obtain
\eqn
        \lim_{N\rightarrow0}\sum_{n,n'=0}^N
        \sum_{\stackrel{\pi\in\Pi_{n, n'}}
        {\pi\;{\rm simple}}}\ampl_{\hat J_\eta,main}[\pi] 
        &=&
        \sum_{m=0}^\infty \lambda^{2m}
        \int dv_0 d\xi
        \overline{\hat J_\eta( \xi,v_0)}
        \nonumber\\
        &\times&\int \Big[ \prod_{j=0}^m
        ds_j\Big]_t \Big[ \prod_{j=0}^m
        d\tilde s_j\Big]_t\int_{(\Tor^3)^m} dv_1\cdots dv_m
        \hat W_{\phi_0^{(\eta)} }(\xi  , v_m)
        \nonumber\\
        &\times&
        e^{2t\lambda^2\Im[\Th (v_0,\e)]}
        e^{-i\sum_{i=0}^m\big( s_i \en(v_i+\xif)+ \tilde s_i \en(v_i-\xif)\big) } \;,
        \label{eqn4-1-1}
\eeqn
where
\eqnn
        \hat W_{\phi_0^{(\eta)} }(\xi , v ) &=&
        \overline{\hat\phi_0^{(\eta)}(v - \xif )}
        \hat\phi_0^{(\eta)}(v + \xif ) \; ,
\eeqnn
and
\eqnn
        \Im[\Th (v_0,\e)]&=&\frac{1}{2}
        \big[\Th (v_0,\e)-\Th(v_0,-\e)\big] \;.
\eeqnn
To derive the macroscopic scaling and weak disorder limit, 
we introduce the new time variables
\eqnn
        a_j := \frac{s_j + \tilde s_j}{2} \;\;,\;\;
        b_j := \frac{s_j - \tilde s_j}{2} \;,
\eeqnn
with $a_j\geq0$ and $\sum_{j=0}^n a_j=t$, and $b_j\in[-a_j,a_j]$.
so that $ds_jd\tilde s_j=2da_jdb_j$, and
\eqn
        s_i \en(v_i-\xif)-\tilde s_i \en(v_i+\xif)&=&
        a_i \big[\en(v_i+\xif)-  \en(v_i-\xif)\big]\nonumber\\
        &+&  b_i
        \big[\en(v_i-\xif)+  \en(v_i+\xif)\big]\;.
        \label{eqn4-1-5}
\eeqn
Furthermore, we introduce macroscopic variables
\eqnn
        T := \eta  t=\eta\e^{-1}
        \; \; , \; \;
        \tau_j := \eta  a_j
        \; \; ,   \; \;
        \zeta := \eta^{-1}\xi
        \; ,
\eeqnn
where we recall from (~\ref{eta-lamb-cond-1}) that the scaling factor and the disorder 
strength are related by
\eqn
        \eta=\lambda^2 \;.
        \label{eta-lamb-cond-2}
\eeqn
We note that $|\zeta|\leq O(1)$ on the support of $\hat J(\zeta,v)$.

For any finite $\tau_j$,
\eqnn
        w-\lim_{\eta\rightarrow0}\prod_{j=1}^{n} \int_{-\tau_j/\eta}^{\tau_j/\eta} db_j
        \, e^{ 2 i  b_j
        ( \en(v_j)-\en(v_0)+O(\eta) )} 
        =\prod_{j=1}^{n}\pi\delta\big(
        \en(v_j)-\en(v_0)\big) \;,
\eeqnn 
and by the same arguments as in \cite{erdyau}, we obtain
\eqn
        \lim_{\eta\rightarrow0}\Exp\big[\langle\hat J_\eta,\hat
        W_{\phi_t^{(\eta)}}\rangle\big] 
        &=&\sum_{n\geq0} \int_{ (\Tor^3)^{n+1}}  dv_0\cdots dv_n
        e^{2T\Im\Th (v_0)}
        \nonumber\\
        &\times&
        2^n\int \Big[\prod_{j=0}^n d\tau_j  \Big]_T
        \prod_{j=1}^{n} \pi \delta\big(\en(v_j)-\en(v_0)\big)
        \nonumber\\
        &\times&
        \lim_{\eta\rightarrow0}\int_{(2\Tor/\eta)^3}d\zeta
        \overline{\hat J(\zeta,v_0)}
        e^{2\pi i\sum_{j=0}^n\tau_j \zeta \cdot \sin 2\pi v_j }
        \nonumber\\
        &\times&
        \hat W_{\phi_0^{(\eta)}}(\eta\zeta , v_n ) \;,
\eeqn
where 
\eqn
	\Th (v):=\lim_{\e\rightarrow0}\Th (v,\e)\;.
\eeqn
We observe that in the $n$-th term of the sum,
the factor $\eta^{-n}$, which emerges from rescaling $a_i$, has  
eliminated $\lambda^{2n}$, due to (~\ref{eta-lamb-cond-2}).
Moreover, using (~\ref{initcondweaklim}), 
\eqn
        &&\lim_{\eta\rightarrow0}\int_{(2\Tor/\eta)^3}
        d\zeta\overline{\hat J(\zeta,v_0)}
        e^{2\pi i\sum_{j=0}^n\tau_j \zeta \cdot \sin 2\pi v_j }
        \hat W_{\phi_0^{(\eta)}}(\eta\zeta , v_n )\nonumber\\
        &=&\int_{\R^3}dX\overline{ J(X,v_0)}
        F_{0}\Big(X-\sum_{j=0}^n  \tau_j  \sin2\pi v_j, v_n \Big) \;.
\eeqn
Thus, for any test function $J(X,V)$, one obtains
\eqn
        \lim_{\eta\rightarrow0}\lim_{N\rightarrow\infty}
        \Exp\big[\langle J, W^{(\eta)}_{\phi_{\eta^{-1}T, N}^{main}}
        \rangle\big]
        =\langle J,  F_T\rangle\; ,
\eeqn
where $W^{(\eta)}_{\phi_{\eta^{-1}T, N}^{main} }$ is the
rescaled Wigner transform corresponding to $\phi_{\eta^{-1}T, N}^{main}$,
and
\eqnn
        F_T(X,V)&=&e^{2T\sigma(V)}\sum_{n\geq0}
        \int d\tau_0\cdots d\tau_n \delta\Big(\sum_{j=0}^n\tau_j-T\Big)
        \nonumber\\
        &\times&
        \int dV_1\cdots dV_n   \sigma(V,V_1)\cdots
        \sigma(V_{n-1},V_n)
        \nonumber\\
        &\times&
        F_0\Big(X-\sum_{j=0}^n \tau_j \sin 2\pi V_j , V_n \Big)\; ,
\eeqnn
with $V=V_0$. Here,
\eqn
        \sigma(V,U) := 2\pi\delta\big(\en(V)-\en(U)\big)
\eeqn
corresponds to the differential cross-section, while
\eqn
        \sigma(V) :=  \int dU \sigma(V,U)=-2\Im[\Th(V)] \;.
\eeqn
is the total scattering cross section.
The key insight is that $F_T(X,V)$
satisfies the linear Boltzmann equations (~\ref{linB}),
hence this result concludes our proof of Theorem {~\ref{Boltzlimthm}}.
\endprf

\subsubsection*{Acknowledgements}

The author is profoundly grateful to L. Erd\"os, and especially H.-T. Yau,
for their support, guidance, and generosity.
He has benefitted immensely from very numerous discussions with H.-T. Yau,
without whom this work would not have been possible.
He is much indebted to
H. Spohn for his advice, encouragement and support, and
to an anonymous referee for very detailed and helpful comments.
He thanks J. Lukkarinen for helpful comments, and
A. Elgart, B. Schlein
for discussions. This work was supported
in part by a grant from the NYU Research Challenge
Fund Program, and in part by NSF grant DMS-0407644.
It was carried out while the
author was at the Courant Institute,
New York University, as a Courant Instructor.

\newpage

\centerline{\epsffile{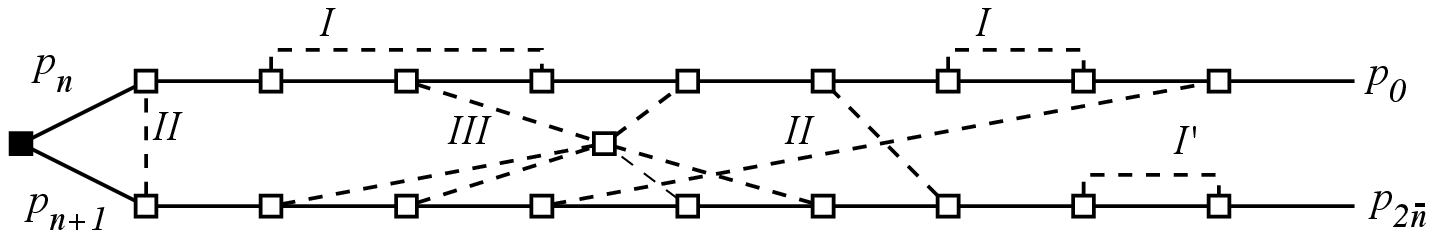} }

\noindent{Figure 1.} A graph containing type I, I', II, and III
contractions.
\\

\centerline{\epsffile{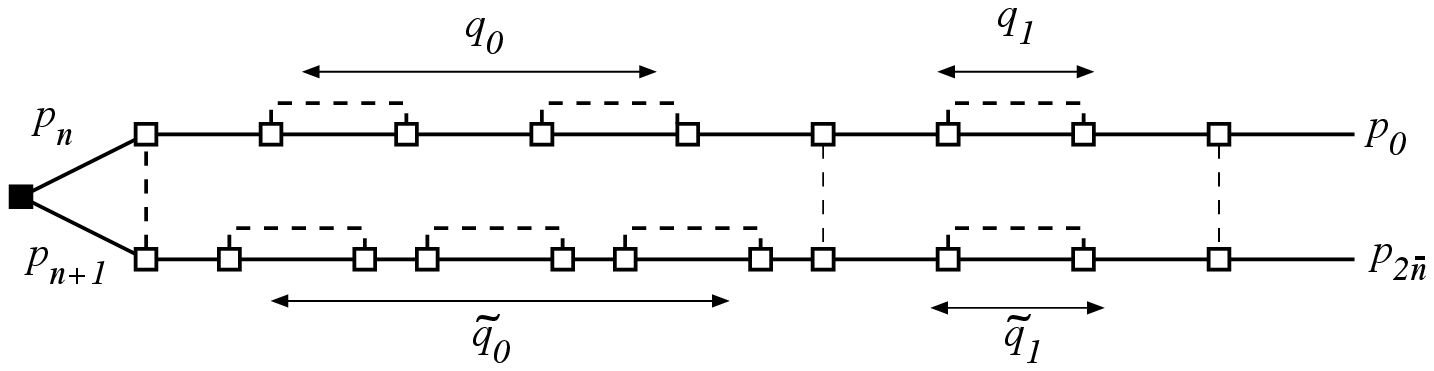} }

\noindent{Figure 2.} A simple pairing contraction graph.
\\

\centerline{\epsffile{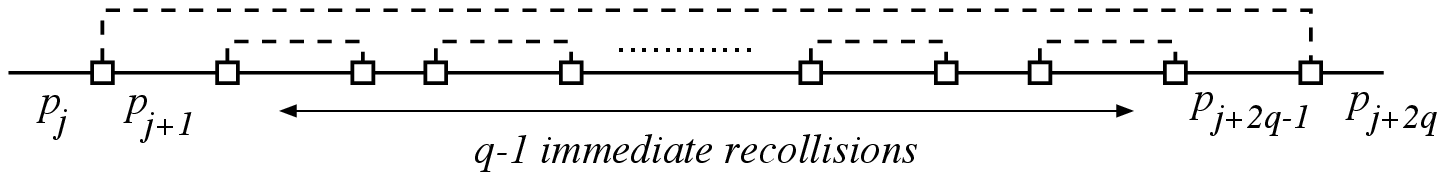} }

\noindent{Figure 3.} A simple nest.
\\


\begin{thebibliography}{99}

\bibitem{cyfrkisi} Cycon, H. L., Froese, R. G., Kirsch, W., Simon, B.,
{\em Schr\"odinger operators}, Springer Verlag (1987).

\bibitem{erd} Erd\"os, L., {\em Linear Boltzmann equation as the scaling limit of
the Schr\"odinger evolution coupled to a phonon bath}, J. Stat. Phys.
107 (5), 1043-1127 (2002).

\bibitem{erdyau} Erd\"os, L., Yau, H.-T., {\em Linear Boltzmann equation
as the weak coupling limit of a random Schr\"odinger equation},
Comm. Pure Appl. Math., Vol. LIII, 667 - 753, (2000).

\bibitem{erdsalmyau} Erd\"os, L., Salmhofer, M., Yau, H.-T., {\em Quantum
diffusion of the random Schr\"odinger evolution in the scaling limit}, preprint
http://xxx.lanl.gov/abs/math-ph/0502025.

\bibitem{mapori1} Magnen, J., Poirot, G., Rivasseau, V., {\em Renormalization group methods
and applications: First results for the weakly coupled Anderson model},  Phys. A 263, no.
1-4, 131-140 (1999).

\bibitem{mapori2} Magnen, J., Poirot, G., Rivasseau, V., {\em Ward-type identities for the
two-dimensional Anderson model at weak disorder}, J. Statist. Phys., 93, no. 1-2, 331-358 (1998).

\bibitem{po} Poirot, G., {\em Mean Green's function of the Anderson model at weak disorder with
an infrared cut-off}, Ann. Inst. H. Poincar\'e Phys. Th\'eor. 70, no. 1, 101-146 (1999).

\bibitem{shscwo} Schlag, W., Shubin, C.,  Wolff, T.,
{\em Frequency concentration and localization lengths for the Anderson model at small
disorders}, J. Anal. Math., 88 (2002).

\bibitem{sp} Spohn, H.,
{\em Derivation of the transport equation for electrons moving through
random impurities}, J. Statist. Phys., 17, no. 6, 385-412 (1977).

\bibitem{st} Stein, E., {\em Harmonic Analysis}, Princeton University Press (1993).



\end{thebibliography}
\end{document}